\documentclass[fleqn,usenatbib]{mnras}

\usepackage{newtxtext,newtxmath}

\usepackage[T1]{fontenc}

\DeclareRobustCommand{\VAN}[3]{#2}
\let\VANthebibliography\thebibliography
\def\thebibliography{\DeclareRobustCommand{\VAN}[3]{##3}\VANthebibliography}

\usepackage{graphicx}	
\usepackage{amsmath}	

\usepackage{amssymb}	
\usepackage{makecell}
\usepackage{float}
\usepackage{multirow}
\usepackage{hyperref}
\usepackage{url}
\usepackage[normalem]{ulem}
\usepackage{lineno}

\usepackage{xspace}

\newcommand{\kev}{\text{\,keV}\xspace}

\newcommand{\muJy}{\text{\,$\mu$Jy}\xspace}
\newcommand{\GHz}{\text{\,GHz}\xspace}
\newcommand{\MHz}{\text{\,MHz}\xspace}

\newcommand{\cps}{\text{\,count\,s$^{-1}$}\xspace}
\newcommand{\xflux}{\text{\,erg\,s$^{-1}$\,cm$^{-2}$}\xspace}

\newcommand{\mins}{\text{\,minute}\xspace}

\newcommand{\dys}{\text{\,days}\xspace}

\newcommand{\counts}{\text{\,counts\,s}$^{-1}$\xspace}

\title[SAX J1810.8-2609]{SAX J1810.8-2609: An Outbursting Neutron Star X-ray Binary with Persistent Spatially Coincident Radio Emission}

\newcommand{\AuthorList}{%
A.~K.~Hughes,$^{1}$\thanks{E-mail: hughes1@ualberta.ca}
G.~R.~Sivakoff,$^{1}$
J.~van~den~Eijnden,$^{2}$
R.~Fender,$^{3}$
J.~C.~A.~Miller-Jones,$^{4}$
and E.~Tremou$^{5}$}

\author[A. K. Hughes et al.]{%
\AuthorList
\\%
$^{1}$Department of Physics, University of Alberta, CCIS 4-181, Edmonton, AB T6G 2E1, Canada \\
$^{2}$Department of Physics, University of Warwick, Coventry CV4 7AL, UK \\
$^{3}$Department of Physics, University of Oxford, Denys Wilkinson Building, Keble Road, Oxford OX1 3RH, UK \\
$^{4}$International Centre for Radio Astronomy Research - Curtin University, Perth, Western Australia
6845, Australia \\ 
$^{5}$National Radio Astronomy Observatory, Socorro, NM 87801, USA \\}

\date{%
Accepted 2023 November 28. Received 2023 November 28; in original form 2023 September 16}

\pubyear{2023}

\begin{document}
\label{firstpage}
\pagerange{\pageref{firstpage}--\pageref{lastpage}}
\maketitle

\begin{abstract}
Here we report on joint X-ray and radio monitoring of the neutron star low-mass X-ray binary SAX J1810.8$-$2609. Our monitoring covered the entirety of its ${\sim}\,5\,$month outburst in 2021, revealing a temporal correlation between its radio and X-ray luminosity and X-ray spectral properties consistent with a `hard-only' outburst. During the outburst, the best-fit radio position shows significant variability, suggesting emission from multiple locations on the sky. Furthermore, our 2023 follow-up observations revealed a persistent, unresolved, steep spectrum radio source ${\sim}\,2\,$years after SAX J1810.8-2609 returned to X-ray quiescence. We investigated potential origins of the persistent emission, which included an unrelated background source, long-lasting jet ejection(s), and SAX J1810 as a transitional millisecond pulsar. While the chance coincidence probability is low (${\lesssim}\,0.16\%$), an unrelated background source remains the most likely scenario. SAX J1810.8$-$2609 goes into outburst every ${\sim}\,5\,$years, so monitoring of the source during its next outburst at higher sensitivities and improved spatial resolutions (e.g., with the Karl G. Jansky Very Large Array or Square Kilometre Array) should be able to identify two components (if the persistent emission originates from a background source). If only one source is observed, this would be strong evidence that the persistent emission is local SAX J1810.8$-$2609, and future monitoring campaigns should focus on understanding the underlying physical mechanisms, as no neutron star X-ray binary has shown a persistent radio signal absent any simultaneous X-ray emission.
\end{abstract}




\section{Introduction}
\label{sec:intro}
Low-mass X-ray binaries (LMXBs) are interacting binary systems that consist of a compact object -- a black hole or a neutron star -- accreting material from a low-mass companion star (${<}\,1\,M_\odot$). The inward-moving accretion flow powers outflows in the form of disk winds and relativistic jets. Many LMXBs are transient systems, spending the majority of their lifetimes in a low-luminosity quiescent state ($L_X \lesssim 10^{32}{\rm\,erg\,s^{-1}}$) before sporadically entering into bright transient outbursts ($L_X > 10^{35}{\rm\,erg\,s^{-1}}$) that last weeks to years \citep[e.g.,][]{2006csxs.book..157M,2006csxs.book...39V}. Since LMXBs rapidly evolve through multiple accretion states during outbursts, LMXBs act as natural laboratories for the study of accretion flows \citep[best measured at X-ray frequencies; e.g.,][]{1999ApJ...519L.159B, 2014MNRAS.443.3270M,2021MNRAS.508..475C} and relativistic jets \citep[best measured at radio through infrared frequencies; e.g.,][]{2002ApJ...573L..35C, 2015MNRAS.450.1745R, 2017MNRAS.469.3141T}. 

The standard accretion state nomenclature (i.e., the \textit{hard} and \textit{soft} accretion states) was developed to describe the different X-ray spectra observed in black hole low-mass X-ray binaries (BHXBs). Moreover, the properties of the relativistic jet(s) are closely correlated with the accretion state \citep[see, ][for detailed reviews]{fender2004,2006ARA&A..44...49R, 2010LNP...794...53B, 2010LNP...794..115F}. In the hard accretion state, the X-ray emission is dominated by high-energy (i.e., hard) X-ray photons comptonized by an optically thin corona. The X-ray spectra are well described by a power law model with a photon index of $\Gamma{\sim}\,1.7$ (where the X-ray flux $f_\text{X}(\nu)\propto\nu^{-\Gamma-1}$). Furthermore, in the hard accretion state, the jet adopts a steady, compact structure. The radio spectrum of the compact jet is the result of a superposition of multiple self-absorbed synchrotron spectra originating from different positions along the jet axis \citep{1979ApJ...232...34B}. At low frequencies, the jet is best described as an optically thick, partially self-absorbed synchrotron spectrum with an inverted or flat spectral index ($\alpha\,{\gtrsim}\,0$; radio flux density $f_R(\nu)\propto\nu^\alpha$) up to a break frequency (often at sub-mm wavelengths). Beyond the break frequency, the jet's spectrum becomes optically thin \citep[$\alpha\,{\sim}\,-0.7$;][]{migliari2010, 2013MNRAS.429..815R, trigo2018}. 

In the hard state, the X-ray ($L_X$) and radio ($L_R = \nu L_{\nu,R}$) luminosities are correlated \citep[henceforth, the $L_R$--$L_X$ relation; ][]{2003MNRAS.344...60G, corbel2013}. After including a scale for the black hole mass, the $L_R$--$L_X$ relation has been extended to include accreting supermassive black holes \citep{2003MNRAS.345.1057M}, thereby spanning 10 orders of magnitude in X-ray luminosity and providing the strongest empirical evidence of the coupling between accretion flows and relativistic jets. Individual BHXBs have exhibited multiple distinct tracks in the $L_R-L_X$ plane \citep[e.g., the `radio-loud' and `radio-quiet' tracks; ][]{2011MNRAS.414..677C,espinasse2018,williams2020,10.1093/mnrasl/slab049} suggesting that the properties of the accretion flow (e.g., geometry and radiative efficiency) may vary significantly in the hard accretion state. Population analyses have both supported \citep[e.g.,][]{gallo2012} and refuted \citep[e.g.,][]{gallo2014,2018MNRAS.478L.132G} the statistically independent existence of multiple tracks, with the more recent studies not finding any robust statistical evidence for separate tracks, suggesting that, instead,  the properties of the ‘radio-loud’ and ‘radio-quiet’ track sources vary significantly from source to source.

Conversely, in the soft accretion state, low-energy (i.e., soft) thermal emission from a multi-color accretion disk dominates the X-ray spectrum. Furthermore, the compact jet is quenched, decreasing in luminosity by $\gtrsim\,$3 orders of magnitude \citep{2011MNRAS.414..677C, 2020MNRAS.498.5772R}. During the hard-to-soft transition, one or more discretized ejection events may be launched. These ejections have been spatially resolved in multiple sources \citep[e.g.,][]{mirabel1994,hjellming1995,hannikainen2001,rushton2017,bright2020}. The radio spectra of the ejecta are characterized by a time-variable self-absorbed synchrotron component \citep[sometimes parameterized as the van der Laan (vdL) model; ][]{1966Natur.211.1131V, 1988ApJ...328..600H, 1995xrbi.nasa..308H}. As ejecta propagate and expand, they become optically thin at (progressively) lower-frequency emission, steepening the radio spectral index to $\alpha\,{\sim}\,-0.7$. Emission from jet ejections can persist from hours to years \citep[e.g.,][]{2019Natur.569..374M,2023ApJ...948L...7B}, and can exhibit variability that is unrelated to any simultaneous evolution of the accretion flow \citep[e.g., through collision with the surrounding interstellar medium;][]{10.1093/mnras/stab864}. As a result, radio observations of jet ejecta must be excluded from the $L_R$--$L_X$ relation.

For neutron star (low-mass) X-ray binaries (NSXBs), their strong intrinsic magnetic fields and solid surfaces complicate the picture. Historically, radio emission was thought to be exclusive to the weakly-magnetic (${<}\,10^{10}\,$G) sub-population, although there have been recent detections of radio emission from strongly-magnetic NSXBs \citep[e.g.,][]{2018Natur.562..233V,vde2021}. The weakly-magnetic NSXBs are most directly analogous to BHXBs; thus, the strongly-magnetic sub-population will not be discussed any further (henceforth, NSXBs only refer to weakly-magnetic NSXBs). NSXBs have two main sub-classes; \textit{atoll} and \textit{Z} sources \citep[named for their tracks in colour-colour diagrams, see][for a review]{2006csxs.book...39V}. Atoll sources tend to be lower luminosity and transient, exhibiting similar hard/soft accretion states as transient BHXBs. In contrast, Z sources are often persistent but show rapid timescale variability. Moreover, although Z sources also transition through multiple accretion states, these states tend to be softer than atoll states \citep{muno2002}. Some NSXBs have shown transitions from Z to atoll behaviour at lower X-ray luminosities \citep[and thus accretion rates, e.g., XTE J1701-462;][]{2007ApJ...656..420H, 2009ApJ...696.1257L}, suggesting that the these may not be unique sub-populations, but instead that Z sources are NSXBs with the largest accretion rates \citep[analogs to the rapidly flaring, semi-persistent BHXBs like GRS 1915+105;][]{migliari2006}.

Transient atoll sources more closely follow the evolution of a `typical' transient BHXBs \citep[see, ][for a review]{migliari2006, 2014MNRAS.443.3270M}. Atoll outbursts exhibit distinct hard (also known as ``extreme island'') and soft (also known as ``banana'') accretion states. Atolls (sometimes) exhibit jet quenching in the soft state. State transition-induced jet ejections have been proposed for atolls, although they have only been observed in Z sources \citep[e.g.,][]{fomalont01,spencer2013}. The `typical' evolution of a transient outburst of an atoll NSXB or BHXB begins with a departure from quiescence through a rapid brightening in the hard state. The source transitions to the soft state following the initial brightening. The system then remains in the soft state for some time (the amount of time varies from system to system) until it begins to dim, eventually returning to the hard state at a lower X-ray luminosity. Once back in the hard state, the system dims until it returns to a quiescent state. However, some systems break this paradigm by exhibiting erratic state transitions \citep[e.g.,][]{2020A&A...634A..94K} or failed (i.e., `hard-only') outbursts \citep[e.g.,][]{Rodriguez2006,10.1093/mnras/stw903,Tarana2018,Stiele2021}. Recent analyses have shown that ${\sim}\,40\%$ of outbursts of BHXBs are thought to be `hard-only' \citep{2016ApJS..222...15T}; this fraction has not been thoroughly explored for NSXBs.

There are several significant differences between the neutron star and black hole X-ray binary sub-populations: (i) NSXBs generally have radio luminosities that are a factor of ${\sim}\,20$ lower than BHXBs at comparable X-ray luminosities \cite[the discrepancy cannot be attributed to the difference in compact object mass;][]{2018MNRAS.478L.132G}; (ii) NSXBs have shown compact jet radio emission in the soft accretion state \citep[e.g.,][]{10.1111/j.1365-2966.2004.07768.x, 2017MNRAS.470.1871G, vde2021}, suggesting the quenching process may not be as extreme as observed in black hole systems or possibly a different jet launching process completely; (iii) all accretion states can have an additional thermal X-ray component \citep[often modeled as a black body component, see][]{2007ApJ...667.1073L} due to emission from the neutron star surface or boundary layer between the accretion disk and surface. Historically, studies of accretion-jet coupling of NSXBs have suffered from their weaker radio emission. Joint X-ray and radio monitoring of NSXBs is critical for understanding the differences between the neutron star and black hole X-ray binary populations and how the presence (or absence) of an event horizon, ergosphere, or solid surface affects the connection between the accretion flow and relativistic jet. In 2021, the NSXB SAX J1810.8$-$2609 exhibited a multi-month outburst that was detected in both X-ray and radio frequencies, allowing for a comprehensive monitoring campaign. 

\subsection{SAX J1810.8$-$2609}
SAX J1810.8$-$2609 (henceforth SAX J1810) is a NSXB that was initially discovered in 1998 by the wide-field X-ray cameras aboard the  \textit{BeppoSAX} satellite \citep{1998IAUC.6838....1U}. Since its discovery, there have been four subsequent (detected) outbursts that occurred in 2007 \citep{2007ATel.1175....1D}, 2012 \citep{2013IAUS..291..141D}, 2018 \citep{2018ATel11593....1N}, and 2021 \citep{2021ATel14649....1I}. A Type I X-ray burst \citep[i.e., the runaway thermonuclear detonation of a hot-dense surface layer of accreted matter, see][for a review]{2021ASSL..461..209G} revealed the presence of a solid surface, identifying the accreting object as a neutron star \citep{2000ApJ...536..891N}. Furthermore, X-ray modelling of the burst showed a clear signature of photospheric radius expansion (PRE), where the burst luminosity exceeds the local Eddington limit causing a radial expansion of the neutron star photosphere. The PRE X-ray burst was used to estimate the source distance of $4.9\,{\pm}\,0.3\,$kpc \citep[see, ][for a review of PRE bursts as standard candles]{2003A&A...399..663K}. However, we note that the quoted distance error is purely statistical, as it does not take into consideration any systematic effects, such as the potential for the neutron star to deviate from the assumed mass of $1.4M_\odot$ or the potential for accreting elements besides hydrogen. Therefore, the error on the distance is likely an underestimation. An analysis of multiple Type I X-ray bursts detected during the 2007 outburst showed timing signals consistent with a neutron star spin frequency of 531.8$\,$Hz \citep{2018ApJ...862L...4B}. These `millisecond burst oscillations' are thought to be caused by anisotropic X-ray emission \citep[i.e., `hot spots';][]{watts2012} and allow for the determination of the neutron star spin frequency without the need for consistent pulsations. 

The source has not been classified as an atoll or Z source; instead, it has adopted the broader label of neutron star `soft X-ray transient', which encompasses both sub-classes. However, given its moderate peak X-ray luminosity ($L_X{\leq}\,4\times10^{36}{\rm\,erg\,s^{-1}}$) and transient behaviour, it is likely to be an atoll source. The majority of Z sources are persistent and bright, with maximum X-ray luminosities reaching appreciable fractions of the Eddington limit (e.g., $L_X{\sim}\,2\times10^{38}{\rm\,erg\,s^{-1}}$). 

On 2021 May 13 (MJD 59347), the gas slit camera (GSC) aboard The Monitor of All-sky X-ray Image \citep[i.e, MAXI;][]{2009PASJ...61..999M} satellite detected the X-ray brightening of SAX J1810 as it entered its fifth recorded outburst \citep{2021ATel14649....1I}. Following the X-ray detection, radio observations with the MeerKAT radio telescope on 2021 May 21 (MJD 59356) revealed a spatially coincident radio source, constituting the first radio detection of this source \citep{2021ATel.14659....1M}. Here we present our multi-instrument radio/X-ray monitoring campaign of SAX J1810. Our monitoring includes the 2021 outburst and 2023 follow-up that revealed the existence of a spatially coincident, persistent steep spectrum radio source. The remainder of this paper is structured as follows: in Section \ref{sec:obs}, we introduce our observation and analysis procedure, while in Sections \ref{sec:results} and \ref{sec:discussion}, we present and discuss our results. Finally, we summarize our findings in Section \ref{sec:conclusions}.

\section{Observations and Data Analysis}
\label{sec:obs}
\subsection{MeerKAT}
\label{sec:MeerKAT}
\subsubsection{Weekly Monitoring}
\label{sec:MeerKAT_Weekly_Monitoring}
We observed SAX J1810 with MeerKAT \citep[a radio interferometer;][]{camilo2018} as a part of the large survey project ThunderKAT \citep{2016mks..confE..13F}. We began a weekly monitoring campaign on 2021 May 22 (MJD 59356), nine days after the outburst's initial detection, and continued until 2021 October 23 (MJD 59508) for a total of 21 observations. Each observation consisted of a single scan of 15 minutes on-source flanked by two 2-minute scans of a nearby gain calibrator (J1830-3602). Each epoch also included a 5-minute scan of PKS B1934-638 (J1939-6342) for flux and bandpass calibration. In addition to the weekly monitoring, we observed two deep (1-hour) epochs on 2023 May 22 (MJD 60086) and 2023 August 16 (MJD 60172) when the source was in (X-ray) quiescence. The deep epochs followed the same observing strategy, except the source monitoring was broken into two 30-minute scans. All MeerKAT observations used the L-band receiver, with a central frequency of 1.3\GHz, and a total (un-flagged) bandwidth of 856MHz split evenly into 32768 frequency channels. To decrease the size of each data set, we averaged together every 32 channels (resulting in 1024 total channels) before data reduction and imaging. This averaging will not affect our final results as we are focused on radio continuum emission (as opposed to spectral lines). 

We performed flagging, calibration, and imaging using a modified version of the semi-automated routine \textsc{OxKAT}\footnote{Found at: \url{https://github.com/IanHeywood/oxkat}} \citep{2020ascl.soft09003H}, which breaks the process into three steps. Here we will briefly outline the workflow and direct readers to \citet{2022MNRAS.509.2150H} for a more comprehensive description. The first step (1GC) uses \textsc{casa} \citep[v5.6;][]{2007ASPC..376..127M} to remove data corrupted by radio frequency interference (RFI). After removing RFI, the data is corrected with standard calibration solutions (i.e., flux density, bandpass, and complex gain). The second step (FLAG) applies a second round of flagging using \texttt{tricolor} \citep{2022ASPC..532..541H} before creating a preliminary image of the source field using \textsc{wsclean} \citep[v2.9;][]{2014MNRAS.444..606O}. This preliminary image is then used to create an imaging mask. The final step (2GC) begins with a masked deconvolution before using the model image for direction-independent (DI) self-calibration with \textsc{CubiCal} \citep{2018MNRAS.478.2399K}. Following self-cal, the pipeline ends with a second round of masked deconvolution using the DI self-calibrated visibilities. We adopted the 2GC images as our final data products. We maximize our sensitivity by weighting each image with a Briggs' robustness of 0 \citep{1995AAS...18711202B}\footnote{MeerKAT's synthesized beam becomes significantly non-Gaussian for robustness weightings ${>}\,0$, inhibiting accurate deconvolution and raising the image-plane rms noise.}. We note that \textsc{OxKAT} has the functionality to solve for direction-dependant (DD) self-calibration solutions if needed (i.e., the 3GC step). However, for SAX J1810, DI self-calibration was sufficient, and thus we omitted the 3GC step.   

We measured the source properties in each epoch using the \textsc{casa} task \texttt{imfit}, fitting an elliptical Gaussian component in a small sub-region around the source to measure the position and flux density. As the source was unresolved, we set the component shape to be the synthesized beam of each image. We quantified the (1$\sigma$) uncertainty on the flux measurement using the local root-mean-square (rms) noise. We extracted the rms from an annular region for each epoch using the \textsc{casa} task \texttt{imstat}. Each annulus was centered on the position of the Gaussian component. We fixed the inner radius as the major axis of the synthesized beam and scaled the outer radius such that the annular area comprises the area of 100 synthesized beams. We quantified astrometric errors using the method detailed in Appendix \ref{sec:appendix_radio_astronomy}.

\subsection{Very Large Array}
\label{sec:VLA}
We were approved for a single director's discretionary time observation (Project Code: 23A--417) with the Very Large Array (VLA) as a follow-up of our initial 2023 MeerKAT observation. SAX J1810 was observed on 2023 July 17 (MJD 60142) in the 2--4\GHz (S-band) and 4--8\GHz bands (C-band). For S-band, the observations used the 8-bit sampler comprised of two base-bands, with eight spectral windows of sixty-four 2\MHz channels each, giving a total (unflagged) bandwidth of 2.048\GHz. The 3-bit sampler was used for C-band, which has four base-bands, and thus a 4.096\GHz bandwidth. In each band, we included a single 1-minute scan of the flux calibrator (3C286). For source monitoring the array cycled between SAX J1810, observed for ${\sim}\,8\,$minutes per cycle in S-band and ${\sim}\,5\,$minutes in C-band. Each source scan is flanked by ${\sim}\,1\,$minute observations of a nearby gain calibrator (J1820$-$2528). The total time on source was ${\sim}\,16\,$minutes in both bands. We performed flagging, calibration, and imaging using the most recent release of the \textsc{casa} VLA pipeline (v6.4). We imaged the source using \textsc{wsclean} but did not detect the source in either band. As a result, we extract the rms noise from each image to place ($3\sigma$) upper limits on the flux density. We used a circular extraction region (with an area equal to 100 synthesized beams) centered on the archival position of SAX J1810 to measure the rms. The radio flux densities from both MeerKAT and the VLA are presented in Table \ref{tab:radio_monitoring}

\subsection{\textit{Swift}-XRT}
\subsubsection{Weekly Monitoring}
\label{sec:Swift_Weekly_Monitoring}
 We monitored SAX J1810 with the X-ray telescope \citep[XRT;][]{2005SSRv..120..165B} aboard the Neil Gehrels \textit{Swift} Observatory \citep{2004ApJ...611.1005G}, capturing the quasi-simultaneous evolution of the X-ray flux (i.e., within $\sim\,$3\dys of a MeerKAT observation). During the outburst, we observed 21 epochs (target ID: 32459) between 2021 May 20 (MJD 59364) and 2021 November 6 (MJD 59524) at an approximately weekly cadence. To accompany our deep MeerKAT epochs, we were approved for two Target-of-Opportunity observations on 2023 May 25 (MJD 60089) and 2023 August 16 (MJD 60172). During the initial stages of the outburst, we monitored the source in Windowed Timing (WT) mode, where SAX J1810 exhibited a maximum count rate of ${\sim}\,20{\rm\,count\,s^{-1}}$ during the first epoch. We transitioned to Photon Counting (PC) mode when the sources count rate decayed to $\lesssim\,$1 count s$^{-1}$ on 2021 October 9 (MJD 59496), although there was a single intermittent PC epoch on 2021 September 5 (MJD 59462). 

 We used the Python API version of the \textit{Swift}-XRT pipeline, \texttt{swifttools} \citep{2007A&A...469..379E,2009MNRAS.397.1177E}, to extract the source and background spectra for all epochs except 2021 August 7 (MJD 59433), where the source exhibited a Type I X-ray burst (see section \ref{sec:methods_xray_burst}). We used the HEASOFT package (version 6.25) for our spectral analysis. For observations that had a sufficiently large number of counts (i.e., MJD 59364--59496), we used a modified \texttt{grppha} script to bin the spectra on 25-count intervals and performed spectral fitting using $\chi^2$ statistics. Towards the end of our 2021 monitoring (i.e., the MJD 59504 and 59511), we used Cash statistics \citep[i.e., \texttt{cstat};][]{1979ApJ...228..939C} with single-count binning intervals, due to the small number of counts collected in each observation. The final two epochs of the 2021 monitoring (MJD 59518 and 59524) and the late-time follow-up (MJD 60089 and 60172) were non-detections and thus were omitted from the spectral fitting routine. 

Using \textsc{xspec} \citep{1996ASPC..101...17A}, we performed our spectral fitting twice, once for the 0.5--10 keV energy range and again for 1--10 keV. As expected, changing the energy range had a negligible effect on the best-fit spectral parameters. We modelled the spectra using an absorbed power law model with an added blackbody component; i.e., \texttt{tbabs} $\times$ (\texttt{pegpwrlw} $+$ \texttt{bbody}), where \texttt{tbabs} models the interstellar absorption using an equivalent hydrogen column density ($N_H$) following the abundances from \citet{2000ApJ...542..914W}. The power law accounts for the X-ray emission from the dominant component (i.e., the hard X-ray corona), and the blackbody accounts for any excess soft X-ray emission from a faint accretion disk, neutron star surface, or boundary layer. Initially, we fit each spectrum individually, allowing $N_H$ to vary epoch by epoch. We then adopted the single epoch fitting as our starting parameters, linking the $N_H$ values across all epochs and fitting the spectra simultaneously, resulting in a single time-independent value of $N_H$. When calculating the degrees of freedom, we treated the linked $N_H$ as frozen (i.e., each spectrum has four free parameters). The epochs that utilized Cash statistics were omitted from the fitting procedure detailed above. Instead, we fit each of those spectra with a simple absorbed power law model (i.e., \texttt{tbabs} $\times$ \texttt{pegpwrlw}), fixing $N_H$ to our best-fit value of $3.88\times10^{21}{\rm\,cm^{-2}}$ and the power law photon index ($\Gamma$) to the average value of $1.61$ from the $\chi^2$ fitting. As a result, the X-ray flux was the only free parameter in the Cash statistic modelling. The \textit{Swift}-XRT monitoring and spectral parameters during the 2021 outburst are presented in Table \ref{tab:swiftxrt_params}. The quoted uncertainties on the X-ray parameters represent the standard $90\%$ confidence intervals. 

\subsubsection{Type I X-ray Burst}
\label{sec:methods_xray_burst}
On 2022 August 7 (MJD 59433), SAX J1810 underwent a Type I X-ray burst, and, as a result, we performed manual data reduction on the \textit{Swift}-XRT (WT) observations. First, we ran the task \texttt{xrtpipeline} to produce cleaned event files and exposure maps. Second, using \texttt{barycorr}, we applied the barycentric timing correction. Lastly, we extracted source and background spectra by using \texttt{xselect}. For the pre-burst times, we used a circular source extraction region with a radius of 30 pixels (1 pixel = 2.36 arcsec) and an annular background extraction region with an inner radius of 70 pixels and an outer radius of 130 pixels. The pre-burst spectrum was then processed using $\chi^2$ statistics and the routine mentioned in \ref{sec:Swift_Weekly_Monitoring}. 

During the burst, we broke the event file into multiple time bins to analyze the time evolution of the spectral parameters. Due to high count rates during the burst (i.e., maximum count rates ${\gtrsim}\,$400\cps), the observations are affected by systematic effects caused by photon pile-up. As a result, we used an annular source extraction region with an inner (exclusionary) radius that increases with an increasing count rate (ranging from 0 to 3 pixels). Following the \textit{Swift}-XRT pipeline procedure \citep[see, ][]{2007A&A...469..379E,2009MNRAS.397.1177E}, we choose inner radii that reduce the maximum count rate in a given time bin to $<$\,150\cps. The time ranges were chosen so each bin has ${\gtrsim}\,300$ counts corresponding to 21 bins across the 1.5\mins burst. To model the burst parameters in \textsc{xspec} we added a second blackbody component to the pre-burst spectrum, fixing the pre-burst parameters, thereby allowing only the second blackbody to vary. We used the \texttt{bbodyrad} model to directly fit for the normalized radius (i.e., size of the blackbody) and temperature before using the \textsc{xspec} convolution model \texttt{cflux} to calculate the flux. 

For the timing analysis, we extracted two light curves. The first light curve was binned on 1$\,$s intervals and was used to model the decay timescales of the burst. We extracted an initial light curve using the circular extraction region. For any time bins with a count rate $>$\,150\cps, we replaced their count rates with the count rate measured by the annular region with a 3-pixel exclusionary inner radius. We corrected for background and annular extraction region effects with \texttt{lcmath} and \texttt{xrtlccorr}, respectively. Following the prescription outlined in \citep{2020ApJS..249...32G} we fit an exponential decay function, 
\begin{align}
    R(t) = Ae^{-\frac{t}{\tau}} + R_0,
\end{align}
where $t$ is the time after the burst maximum, $R(t)$ is the count rate at a given $t$, $R_0$ is the constant background rate, $\tau$ is the $e$-folding decay time, and $A$ is the peak count rate of the bursting component (excluding the contribution from a constant background). We fit for $\tau$, $R_0$, and $A$ with a Markov-Chain Monte Carlo (MCMC) routine using Python's \textsc{emcee} package \citep{2010CAMCS...5...65G,2013PASP..125..306F}, assuming the sampled count rates were independently distributed normal random variables. The number of (sampling) walkers was fixed at five times the number of dimensions (i.e., 15). We chose three flat priors to ensure an unbiased analysis. To ensure convergence, we manually inspected the walkers over many autocorrelation times. Additionally, we analyzed the evolution of the autocorrelation time as a function of the number of MCMC steps following the routine outlined in the \textsc{emcee} documentation\footnote{The documentation can be found here: \url{https://emcee.readthedocs.io/en/stable/tutorials/autocorr/}}. 

The second light curve was extracted using the circular extraction region and binned on 1.8$\,$ms intervals (the minimum bin size possible for WT mode). We used the short timescale light curve to search for millisecond burst oscillations. Given the short timescale binning, no corrections were applied to the 1.8$\,$ms light curves. Appendix \ref{sec:appendix_xray_burst} presents the X-ray burst properties.

\subsection{The WATCHDOG Pipeline}
\label{sec:WATCHDOG}
We calculated the X-ray hardness ratio (HR) using a modified version of the pipeline developed for the Whole-sky Alberta Time-resolved Comprehensive black hole Database Of the Galaxy \citep[WATCHDOG; see][for a comprehensive description of the pipeline]{2016ApJS..222...15T}. The hardness ratio is the ratio between the number of counts in the hard and soft X-ray bands. We used the MAXI/GSC 4--10 keV band as the soft band and 15--50$\,$keV observations from the Burst Alert Telescope \citep[BAT;][]{2005SSRv..120..143B} aboard \textit{Swift} as the hard band. Both sets of observations are publicly available\footnote{MAXI/GSC: \url{http://maxi.riken.jp} \\ \textit{Swift}-BAT: \url{https://swift.gsfc.nasa.gov/results/transients/}}. We modified the pipeline to average daily observations, ensuring the hard X-ray band had a ${\geq}3\sigma$ detection. For data where the soft X-ray band detection significance was ${<}\,3\sigma$, we replaced the measured count rate with $3 \times $ the noise value to estimate a conservative $3\sigma$ lower limit. The source appears to have undergone a hard-only outburst, and, as a result, to get meaningful constraints, we needed to measure either a lower limit or detection on the hardness ratio. No further modifications were applied to the WATCHDOG pipeline.

WATCHDOG defined empirical HR limits that corresponded to the different X-ray states: (i) $C_\text{hard} = 0.3204$; and (ii) $C_\text{soft} = 0.2846$. A hardness ratio is considered consistent with the hard (soft) state if its lower (upper) error bars are above (below) the $C_\text{hard}$ ($C_\text{soft}$) limits. If neither criterion is met, the source is classified as being in an intermediate state. We note that the values of $C_\text{hard}$/$C_\text{soft}$ were calculated for BHXBs; in Section \ref{sec:hard_only_outburst}, we investigate whether it is valid to apply the same standard NSXBs. 

\section{Results}
\label{sec:results}
\subsection{Radio Position}
\begin{figure}
    \centering
    \includegraphics[width=0.9\linewidth]{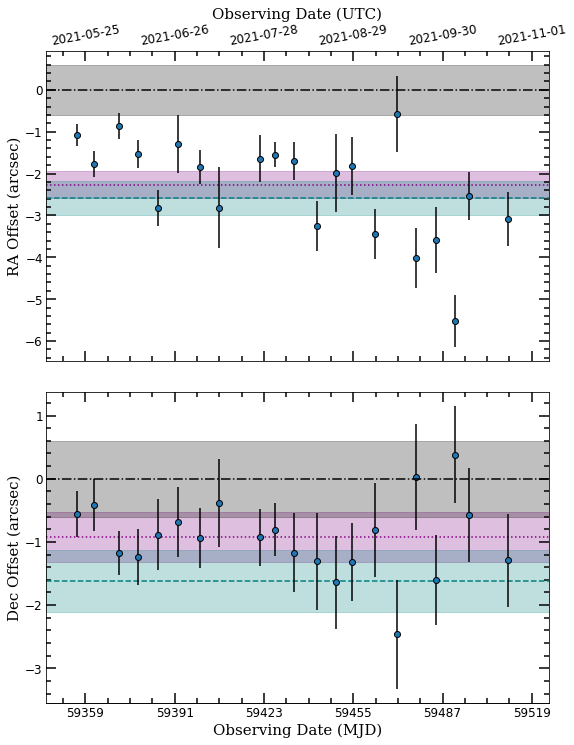}
    \caption{The right ascension (top panel) and declination (bottom panel) offsets for the best-fit SAX J1810 positions. The filled blue circles are the offsets of the source. The purple dotted line and cyan dashed line are the 2023 May 22 and 2023 August 13 offsets, respectively. The dashed-dotted black line is the archival X-ray position from \citet{2004MNRAS.349...94J}, and the grey shaded area is the error on the archival position (${\pm}\,0.6^{\prime\prime}$). Note the clear offset and temporal variability in the right ascension of the source.}  
    \label{fig:sax_astro}
\end{figure}

In Fig.~\ref{fig:sax_astro}, we show the offset in right ascension and declination between the MeerKAT position and the archival X-ray position of 18h10m44.47s $-$26$^\circ$09$^\prime$01.2$^{\prime\prime}$ from \citep{2004MNRAS.349...94J}. The average radio position is 18h10m44.34s $-$26$^\circ$09$^\prime$02.1$^{\prime\prime}$ ($\pm0.1^{\prime\prime}$). The per-epoch declinations are consistent with the average radio position with a reduced $\chi^2=0.75$ (22 degrees of freedom), although the average radio position is offset by ${\sim}\,1^{\prime\prime}$ from the X-ray position. In contrast, the right ascensions show significantly larger offsets ranging from ${\sim}\,$1--$5^{\prime\prime}$. Moreover,  the measured right ascensions show temporal variability. Adopting the weighted mean offset in right ascension as a model and computing the reduced $\chi^2$ results in a value of  $\chi^2=4.4$ (22 degrees of freedom), suggesting that the variability is not the result of stochastic error fluctuations. We tested the right ascension offsets against a linearly increasing model (i.e., ballistic motion), which resulted in a negligible improvement in the reduced $\chi^2$ (4.2; 21 degrees of freedom), and thus, we found no evidence of ballistic motion.

\begin{figure*}
    \centering
    \includegraphics[width=0.9\linewidth]{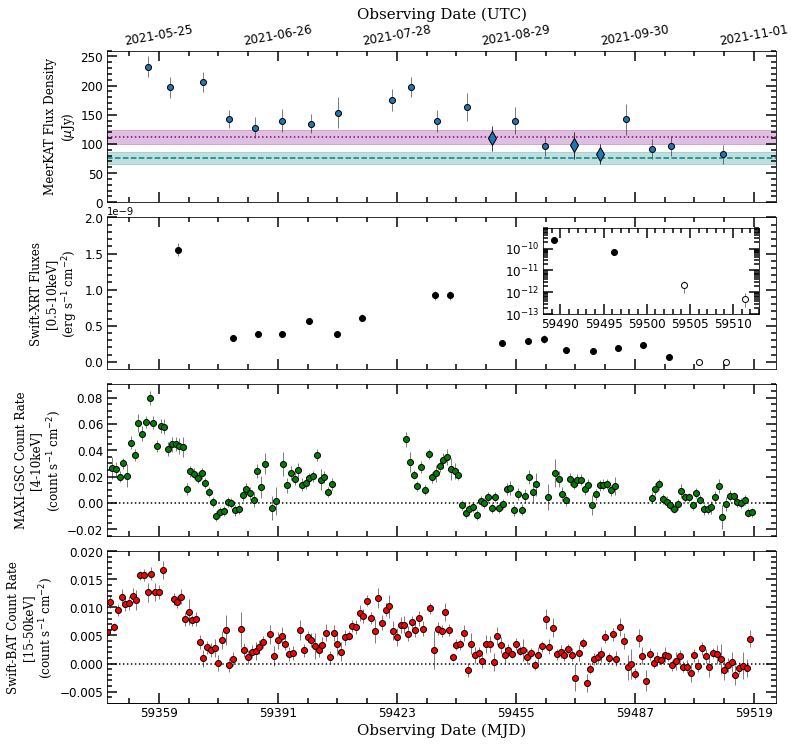}
    \caption{Multi-instrument light curves of the 2021 outburst of SAX J1810. The top panel is the MeerKAT 1.3\GHz radio light curves showing both ${\geq}\,5\sigma$ (blue circles) and $4\text{--}5\sigma$ (blue diamonds) detections during the 2021 outburst. The horizontal lines show the 2023 May 22 (purple dotted) and 2023 August 13 (cyan dashed) flux densities. The second panel is the \textit{Swift}-XRT (0.5--10.0$\,$keV) light curves. The filled and open circles correspond to the epochs fit with $\chi^2$ and Cash statistics, respectively. The bottom two panels show the MAXI/GSC (third panel) and \textit{Swift}-BAT (bottom panel) daily-binned light curves. All four instruments show a common temporal evolution characteristic of the correlation between radio and X-ray emission in the hard accretion state.}
    \label{fig:Light_Curves}
\end{figure*}

\subsection{Outburst Light Curves}
\label{sec:light_curves}
In Fig.~\ref{fig:Light_Curves} we show the MeerKAT (1.3\GHz; top panel), \textit{Swift}-XRT (0.5-10\kev; second panel), MAXI/GSC (4-10\kev; third panel), and \textit{Swift}-BAT (15-50\kev; bottom panel) outburst light curves. For our MeerKAT observations, 18 (out of 21) epochs were ${\geq}\,5\sigma$ detections (blue circles). The remaining three epochs (blue diamonds) do not meet the typical reporting threshold of $5\sigma$, with detection significance of ${\sim}\,4.3$--$4.9\sigma$. Given the spatial coincidences between the low (${<}\,5\sigma$) and high-significance detections (${\geq}\,5\sigma$), it is likely that we are detecting a source in all of our MeerKAT observations. For the \textit{Swift}-XRT light curve, we adopted the total fluxes from our spectral fits using the joint power law and blackbody model components (filled black circles). The last two data points (open black circles) correspond to the epochs where the source was too faint for multi-component spectral modelling; instead, we fit the source with a single power law component. The \textit{Swift}-BAT and MAXI/GSC light curves display the data at a daily binning frequency.

The observed flux of SAX J1810 displays a common temporal evolution across all observing frequencies. At early times ($\sim$ MJD 59340--59370), all four instruments recorded the brightest signal of the outburst. Following the maxima, the source flux began decreasing, showing a rebrightening between $\sim$ MJD 59410 and 59440, before the source flux continued to decrease, returning to X-ray quiescence and plateauing at ${\sim}\,90\,\mu$Jy in the radio. We find no evidence for additional intra-observation variability beyond the Type I outburst discussed in this paper. 

Although the radio and X-ray light curves share a similar evolution in time, the magnitude of the variability is significantly different. In radio, the source exhibits modest variability with a maximum (${\sim}\,230$\muJy) and minimum (${\sim}\,80$\muJy) flux density separated by a factor of only ${\sim}\,3$. In contrast, when only considering the epochs with multi-component spectral modelling, the \textit{Swift}-XRT fluxes show a factor of $\sim\,$20 in variability, with a maximum and minimum flux of ${\sim}\,1.6\times10^{-9}$\ and $6.8\times10^{-11}$\xflux, respectively. Including the final two \textit{Swift}-XRT epochs during the source's return to quiescence, the minimum flux is ${\sim}\,5\times10^{-13}$\xflux, which corresponds to a factor of ${\sim}\,2000$ decrease from the maximum. The plateauing radio emission at MJD 59463 (and beyond) is consistent with a spatially coincident, persistent radio source (see Section \ref{sec:persistent_emission}).

\subsection{X-ray Spectra}
\label{sec:xray_spectra}
\begin{figure*}
    \centering
    \includegraphics[width=0.89\linewidth]{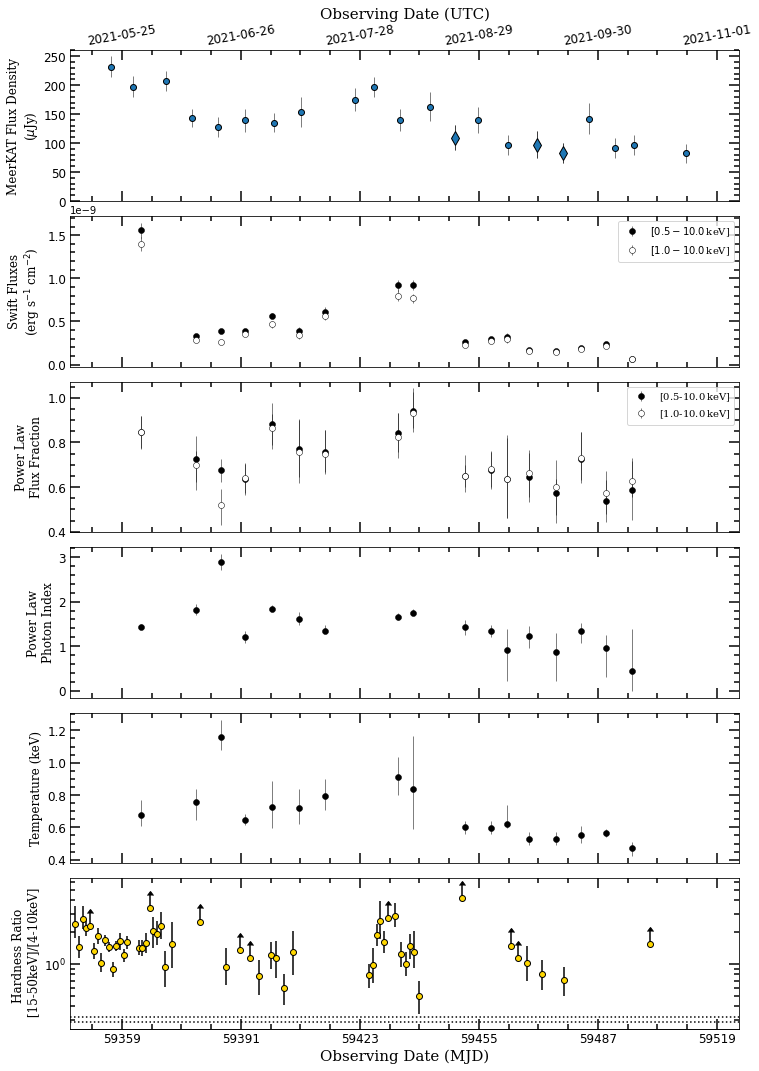}
    \caption{A summary of the spectral properties of SAX J1810. The top panel shows the MeerKAT radio flux density. The next two panels show the \textit{Swift}-XRT X-ray flux (second panel) and the power law flux fraction (third panel) in the 0.5--10.0$\,$keV (filled circles) and 1.0--10.0$\,$keV (open circles) energy bands. The fourth panel shows the temperature of the black body component, and the fifth shows the power law photon index. The bottom panel shows the hardness ratio between the MAXI/GSC (4.0--10.0$\,$keV) and \textit{Swift}-BAT (15.0--50.0$\,$keV) energy bands. The upper ($C_\text{hard}$) and lower ($C_\text{soft}$) dotted lines show the empirically defined state boundaries from WATCHDOG \citep{2016ApJS..222...15T}. These spectral properties are characteristic of the hard accretion state}
    \label{fig:xray_spectra}
\end{figure*}

The X-ray modelling parameters are shown in Fig.~\ref{fig:xray_spectra}. The best fit equivalent hydrogen column density is $N_H = 3.9_{-0.2}^{+0.1}\times10^{21}{\rm\,cm}^{-2}$. The Colden: Galactic Neutral Hydrogen Density Calculator\footnote{The webtool can be found here: \url{https://cxc.harvard.edu/toolkit/colden.jsp}} estimates a value of $N_H\,{\sim}\,(3.2$--$4.3)\times10^{21}{\rm\,cm}^{-2}$ along the SAX J1810 line of sight (depending on the choice of neutral hydrogen data set --- NRAO or Bell), making the measured $N_H$ consistent with expectation. 


To investigate the relative contributions of each model component, we calculated the \textit{power law flux fraction} (third panel, Fig.~\ref{fig:xray_spectra}); i.e., $F_{X,\text{PL}}/F_{X,\text{tot}}$, where $F_{X,\text{PL}}$ is the X-ray flux of the power law component and $F_{X,\text{tot}}$ is the total X-ray flux of the model. In all epochs, the power law component is dominant with a flux fraction ranging from $\sim$ 0.53 to 0.94 with a (variance-weighted) average of $0.72\pm0.02$. The power law photon index ($\Gamma$; fourth panel, Fig.~\ref{fig:xray_spectra}) shows moderate variability with  $0.4_{-0.43}^{+0.93}\leq\Gamma\leq2.88_{-0.08}^{+0.18}$ and an average value of $1.61\pm0.03$. The average value is typical of comptonized hard state X-ray emission from (black hole) X-ray binaries \citep{2006ARA&A..44...49R}. Moreover, if we exclude the anomalously steep photon index, the maximum photon index becomes $\Gamma = 1.83_{-0.08}^{+0.10}$. The blackbody temperature ($kT$; third panel, Fig.~\ref{fig:xray_spectra}) varied between $0.5_{-0.08}^{+0.18} \leq kT \leq 1.2_{-0.08}^{+0.18}{\rm\,keV}$, with an average blackbody temperate of $kT = 0.60 \pm 0.01{\rm\,keV}$. Black body temperatures $\lesssim 1{\rm\,keV}$ are consistent with past analyses of hard state neutron star X-ray binaries \citep[e.g.,][]{2007ApJ...667.1073L}. The bottom panel of Fig.~\ref{fig:xray_spectra} displays the hardness ratio calculated from the daily \textit{Swift}-BAT and MAXI/GSC light curves. We observe a moderate degree of variability in hardness ratio, with detections ranging from ${\sim}\,0.5$--$2.8$, and an average value of $1.19\,{\pm}\,0.06$. Including the lower limits increases the maximum hardness ratio to ${\sim}\,4$.

\renewcommand{\arraystretch}{1.5}
\begin{table}
    \centering
    \begin{tabular}{cccc}
        \Xhline{1\arrayrulewidth}
         Model Component & $\Gamma$ & $kT\,$(keV) & $F_X\,(10^{-11}\,{\rm erg\,s^{-1}\,cm^{-2}})$ \\
         \Xhline{1\arrayrulewidth}
         \texttt{pegpwrlw} & $1.8_{-1.5}^{+0.7}$ & --- &  $18_{-10}^{+7}$  \\
         \texttt{bbody} & --- & $0.9_{-0.1}^{+0.2}$ &   $9_{-2}^{+2}$    \\
         \texttt{diskbb} &  --- & $0.22_{-0.03}^{+0.03}$ &  $12_{-3}^{+2}$ \\
         \Xhline{1\arrayrulewidth}

    \end{tabular}
    \caption{Three component fit of the \textit{Swift}-XRT observation on MJD 59385. We fixed our best fit value to $N_H = 3.88\times10^{21}{\rm\,cm}^{-2}$, and left all other parameters free. After including the \texttt{diskbb} component, the \texttt{pegpwrlw} becomes the subdominant component, and the fit becomes insensitive to both the flux and the photon index of the power law component. SAX J1810 may have briefly entered a thermal X-ray dominated accretion state before returning to the hard state.}
    \label{tab:3comp}
\end{table}

\begin{figure*}
    \centering
    \includegraphics[width=0.9\linewidth]{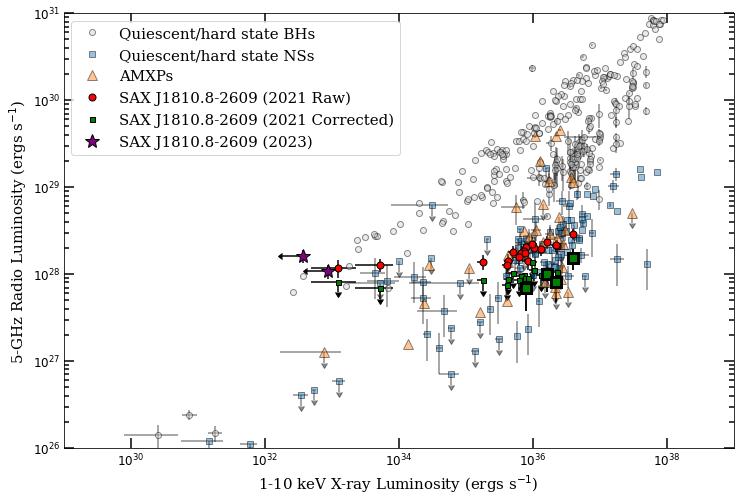}
    \caption{The radio (5\GHz)  and hard state X-ray (1--10$\,$keV) luminosity ($L_R$-$L_X$) relation. The archival values for black holes (grey circles), neutron stars (blue squares), and accreting millisecond X-ray pulsars (orange triangles) were taken from \citet{2022MNRAS.516.2641V}, which is based on the \citet{arash_bahramian_2022_7059313} catalog. The luminosities for SAX J1810 during the 2021 outburst are represented as red circles. The purple stars are the 2023 values if we assume that all radio emission originates from a hard state jet. The late-time plateau in the radio flux density strongly suggested the existence of a second radio source is uncorrelated with the X-ray emission. The green squares show the 2021 outburst values after subtracting $89\mu$Jy from each of the radio flux densities (i.e., the average contribution from the persistent source). Even after subtracting off the persistent source, the two $3\sigma$ radio detections (large green squares) of SAX J1810 remain consistent with the general population of hard-state NSXB jets.}
    \label{fig:lrlx}
\end{figure*}

The largest single epoch evolution occurs on MJD 59385, where the black body temperature reaches its maximum value of ${\sim}\,1.2$\kev, alongside the extreme softening of the power law component ($\Gamma\,{\sim}\,2.9$). During this epoch, the two-component fit had a reduced $\chi^2$ value of ${\sim}\,1.17$ (216.5/186). To investigate whether we were observing a transition to an intermediate or soft state, we added a multi-colour disk to the two-component model; i.e., \texttt{tbabs} $\times$ (\texttt{pegpwrlw} $+$ \texttt{bbody} $+$ \texttt{diskbb}). The inclusion of the third component moderately reduces the $\chi^2$ to ${\sim}\,1.12$ (206.1/184) and decreases both the power law photon index and blackbody temperature to levels consistent with the other epochs (See Table \ref{tab:3comp} for the full model parameters). Moreover, the power law component becomes sub-dominant, suggesting that the source may have briefly transitioned into an intermediate or soft state. The observations on MJD 59413 and 59462 show similarly large reduced $\chi^2$ values of ${\sim}\,1.22$ (237/194) and ${\sim}\,$1.52 (50/33), respectively. As a result, we attempted to fit these spectra with the same three-component model. However, the fitting resulted in a negligible improvement of the $\chi^2$ statistic. We note that, for the latter epochs, both have reduced $\chi^2$ deviations that are consistent (at the ${<}\,3\sigma$ level) with the expected value of 1. Therefore, the poor fits may result from statistical effects rather than a physical change in the X-ray spectrum.

\subsection{Persistent Emission and the ${L_R}$--${L_X}$ relation}
\label{sec:pe_and_lrlx}
Our 2023 follow-up MeerKAT observations revealed a 112$\,{\pm}\,12\,\mu$Jy radio (point) source on 2023 May 22 (MJD 60086) and another 75$\,{\pm}\,11\,\mu$Jy radio source three months later on 2023 August 13 (MJD 60169). The best-fit positions of both 2023 detections are consistent with the 2021 outburst (see Fig.~\ref{fig:sax_astro}). Therefore, we confidently detect a persistent radio source spatially coincident with SAX J1810. We calculated an (intra-band) spectral index of the persistent source using the brighter of the two MeerKAT follow-up observations (MJD 60086). We broke our observations into four evenly spaced sub-bands, ensuring a ${\geq}\,5\sigma$ detection in each sub-band. Applying a simple linear least squares fit, we measured a spectral index of $\alpha=-0.7\,{\pm}\,0.5$. In addition to the large statistical error, we note that intra-band spectral indexes are known to bias towards flatness ($\alpha\,{\sim}\,0$) at detection significances ${\lesssim}\,35\sigma$ \citep{Heywood2016}. Given our source was only detected at ${\sim}\,10\sigma$ and the relatively large error bar, we do not apply any strong physical inference based on this intra-band spectral index

During the last seven epochs of 2021 monitoring (MJD 59463 to 59511) -- after the radio flux density had plateaued -- the average radio flux density is $93\,{\pm}\,7\mu$Jy. This value is consistent with our 2023 observations (at the ${\sim}\,2\sigma$ level), suggesting the persistent emission is, at most, weakly variable with a ${\sim}\,20\%$ excess variance. Combining the late-time 2021 and 2023 observations results in a (weighted) average flux density of $89\,{\pm}\,5\mu$Jy. The quasi-simultaneous \textit{Swift}-XRT follow-up on MJD 60089 and 60172 did not detect any spatially coincident X-ray source in either epoch setting $3\sigma$ upper limits on the 1--10$\,$keV X-ray flux of ${<}\,1.3\times10^{-13}$\xflux and ${<}\,3.0\times10^{-13}$\xflux, respectively. Furthermore, our scheduled VLA follow-up at 3\GHz and 6\GHz, taken between our two MeerKAT observations on 2023 July 17 (MJD 60142), did not detect the source. The 3$\sigma$ upper limits on the 3\GHz and 6\GHz were $30\,\mu$Jy and $18\,\mu$Jy, respectively. Adopting a 1.3\GHz flux density of $78\,\mu$Jy (conservatively assuming a $3\sigma$ drop in flux caused by intrinsic variability), we use the 3\GHz non-detection to calculate a conservative upper limit of $\alpha\,{<}-1.1$. 

Figure \ref{fig:lrlx} presents the $L_R$--$L_X$ relation. The plot includes archival hard state BHXBs (grey circles), hard state NSXBs (blue squares), and accreting millisecond X-ray pulsars (AMXPs; orange triangles). The archival sources were adapted from Fig.~4 of \citet{2022MNRAS.516.2641V}, an updated version of the \cite{arash_bahramian_2022_7059313} catalog. As our \textit{Swift}-XRT and MeerKAT observations were quasi-simultaneous, we applied a one-dimensional linear interpolation to map the radio observations onto the X-ray times for our 2021 observations. We did not apply any interpolation for our 2023 follow-up observations. Instead, we grouped the MeerKAT observations with the nearest \textit{Swift}-XRT follow-up. We present the $L_R$--$L_X$ relation from the 2021 outburst as red circles. Fitting the 2021 results with a simple power law results in a shallow exponent of $\beta=0.09\,{\pm}\,0.03$ (for $L_X\,{\propto}\, L_R^\beta$). If we assume that the 2023 MeerKAT detections originate from a persistent hard state jet (purple stars on Fig.~\ref{fig:lrlx}) and thus should follow the $L_R$--$L_X$ relation, the measured power index becomes an upper limit (due to the X-ray non-detections) adopting a value of $\beta\,{<}\,0.06$. Given that our results strongly suggest the existence of a persistent radio source that is unrelated to the hard state jet of SAX J1810, we present a secondary set of $L_R$--$L_X$ data points (green squares) after subtracting off $93\,\mu$Jy from each of the radio flux densities from our 2021 outburst. Post-subtraction, there are only four epochs (MJD 59364, 59378, 59413, and 59437) that show a ${>}\,3\sigma$ excess flux density when compared to the persistent level. For the rest of the epochs, we set the radio flux density to be $3\times$ the rms noise and displayed them as upper limits. The subtracted values are unconstraining but consistent with the broader population of NSXBs. The implications of SAX J1810 $L_R$--$L_X$ evolution and the origin of the persistent radio source are discussed in Section \ref{sec:persistent_emission}

\section{Discussion}
We monitored the NSXB SAX J1810 during its 2021 outburst. The X-ray and radio properties suggest that the source underwent a `hard-only' outburst, never fully transitioning to a soft accretion state. Moreover, the late-time plateau of radio flux density in 2021, combined with our follow-up in 2023, suggests the existence of a persistent radio source. In the following subsections, we present the evidence of a `hard-only' outburst and discuss the possible origins of the persistent radio emission. 
\label{sec:discussion}
\subsection{Hard-Only Outburst}
\label{sec:hard_only_outburst}

\begin{figure}
    \centering
    \includegraphics[width=0.9\linewidth]{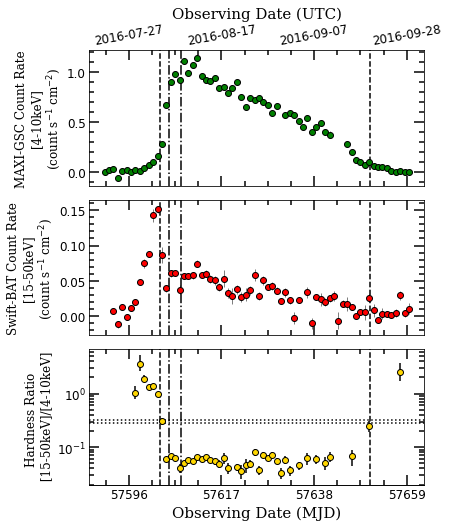}
    \caption{The X-ray evolution of the NSXB Aql X-1's 2016 outburst as seen by MAXI/GSC (top panel), \textit{Swift-BAT} (middle panel), and the hardness ratio between the two instruments (third panel). The horizontal dotted lines adopt the same definition as BHXBs in Fig.~\ref{fig:xray_spectra}. The vertical dashed lines and dashed-dotted lines show the times when the source was independently identified to be in the soft and hard accretion states, respectively \citep{2018A&A...611A..87T}. Both soft accretion states occur at an $\textrm{HR}\,{\sim}\,0.05$, well below the empirically defined transition values. This suggests one can use the BHXB transition hardness ratio to conservatively estimate if a NSXB undergoes a `hard-only' outburst.}
    \label{fig:aqlx1}
\end{figure}

Our observations suggest that SAX J1810 exhibited a `hard-only' outburst in 2021. We justify this claim with three points of evidence:
\begin{enumerate}
    \item The hardness ratio between the \textit{Swift}-BAT and MAXI/GSC observations is above the hard state limit throughout the monitoring. Although the limit was empirically defined using outbursting BHXBs, we expect that the persistent source of thermal X-ray photons (from the neutron star surface or boundary layer) would make all X-ray states softer, thereby decreasing the hard state limit for NSXBs. We investigate this proposition by analyzing the best-studied outbursting (atoll) NSXB, Aql X-1. In Fig.~\ref{fig:aqlx1}, we have plotted a sample light curve of Aql X-1 during its 2016 outburst. The source exhibits a rapid transition of its hardness ratio, with a large fraction of the outburst remaining at a steady value of $\,{\sim}\,0.05$ well below the soft state limit derived for BHXBs. \citet{trigo2018} performed an X-ray spectral analysis of four separate observations; the authors identified that the source was in the hard accretion state on 2016 Aug 3 (MJD 57603) and 2016 Sep 19 (MJD 57650) and in the soft accretion on 2016 Aug 5 (MJD 57605) and 2016 Aug 7 (MJD 57607). The hard and soft state epochs are shown with the dashed and dashed-dotted lines in Fig.~\ref{fig:aqlx1}. As expected, the soft and hard state epochs are temporally consistent with small and large hardness ratios. The final (Sep 19) hard state epoch shows a hardness ratio below the BHXB hard state limit, consistent with our prediction that the thermal photons from neutron stars will lower the hard state limits. We note that other outbursts of Aql X-1 \citep[e.g., the 2009 outburst; ][]{2010ApJ...716L.109M} show a similar `softening` of the hard state limit. Therefore, we are confident that the \textit{Swift}-BAT and MAXI/GSC hardness ratio for SAX J1810 is consistent with hard state emission throughout the 2021 outburst, and our adoption of the WATCHDOG limits is most likely appropriate (if not a conservative approximation). 
    
    \item Our \textit{Swift}-XRT spectral modelling is consistent with hard state emission in nearly all epochs. The X-ray photon indexes ($\Gamma_\text{avg}{\sim}\,1.6$) and low-energy black body temperatures  ($kT_\text{avg}{\sim}\,0.6$) are typical of hard state X-ray emission from an NSXB \citep{2007ApJ...667.1073L}. Moreover, the power law component is the dominant flux component in all epochs (i.e., power law flux fraction ${\geq}\,50\%$). Although some epochs show approximately equal contributions between the blackbody and power law components, the narrow ($0.5-10.0\,$keV) energy range favors the black body component when calculating band limit flux, as the power law component will dominate at higher energies (${\geq}\,10\,$keV). The bolometric X-ray flux is more strongly dominated (${>}\,90\%$) by the power law component than our observations would suggest, consistent with hard state emission. The anomalous epoch (MJD 59385; Table \ref{tab:3comp}) that shows a clear softening of the X-ray spectrum suggests the source may have exhibited a brief deviation from a hard accretion state. Assuming a successful transition to the soft state, and given the cadence of our observations and the bracketed hard sate epochs, the source would have gone through a full cycle (i.e., hard $\rightarrow$ soft $\rightarrow$ hard) in ${\leq}\,14\,$days before remaining in the hard state for the remaining ${\sim}\,120\,$days of outburst \citep[atypical behaviour for an outbursting NSXB, see,][ for a review of outburst timescales]{2014MNRAS.443.3270M}. We find it more likely that the source briefly entered an intermediate state, failed to complete a transition to the soft state, and transitioned back to the hard state. 

    \item The evolution of our radio observations is consistent with the hard state. First, the radio and X-ray light curves show a correlated temporal evolution characteristic of hard state emission. Second, we do not detect any significant jet-quenching. Although radio emission from NSXBs has been observed in the soft state, when both hard and soft state (compact jet) radio emission has been detected, the jet emission is brighter in the hard state \citep[at a fixed X-ray flux, e.g., ][]{2017MNRAS.470.1871G}. Therefore, without a significant increase in the X-ray flux (which was never observed), we would expect a decrease in the radio flux after a transition to the soft state. We recognize that the spatially coincident, persistent radio source contaminates our ability to detect jet-quenching. However, the persistent source can not explain the joint radio--X-ray time evolution, as we would expect the radio flux to drop to the persistent level (${\sim}\,90\,\mu$Jy) without a similar decrease in X-ray flux. Whenever we observed an increasing X-ray flux, we observed a simultaneous increase in the radio flux density. 
\end{enumerate}

Comprehensive monitoring campaigns of future outbursts of SAX J1810 will be critical for confirming whether the source consistently exhibits `hard-only' outbursts or shows a broader outburst phenomenology that sometimes results in successful transitions to the soft state \citep[as observed in some BHXBs, e.g., H1743-322;][]{2011MNRAS.414..677C,williams2020}.

\subsection{The Origin of the Persistent Radio Emission}
\label{sec:persistent_emission}
Our observations strongly support the existence of an unresolved, persistent, steep-spectrum radio source spatially coincident with the position of SAX J1810 (${\pm}\,3^{\prime\prime}$). Considering the source exhibited a `hard-only' outburst in 2021, we expect the radio emission to (partially) originate from a hard state jet (i.e., compact jet). The temporal coincidence between the flares at X-ray and radio frequencies is strong evidence for the existence of a steady jet. Moreover, the persistent source is weakly variable with an average flux density of ${\sim}\,90\mu$Jy. Considering that we have multiple detections at ${\gtrsim}\,200\mu$Jy, we have clearly detected radio emission from the compact jet. 

However, a hard state jet associated with SAX J1810 cannot be the source of the persistent radio emission. Hard state jets are stationary and, therefore, would not exhibit the proper motion that we have observed (Fig.~\ref{fig:sax_astro}) Moreover, the locations of its luminosities on the $L_R$--$L_X$ plane (red circles Fig.~\ref{fig:lrlx}) are inconsistent with a hard state jet. At early times and high X-ray luminosities, the radio/X-ray luminosities are positively correlated, as expected from a compact, steady jet. Towards the end of the outburst (at $L_X\,{\lesssim}\,5\times10^{35}{\rm\,ergs\,s^{-1}}$), there is a clear flattening of the correlation resulting in a $\beta{<}\,0.06$ due to the radio luminosity remaining approximately constant while the X-ray luminosity decreased by over three orders of magnitude. The 2023 follow-up, in particular, would make SAX J1810 exceptionally radio-loud for a NSXB, consistent with the population of BHXBs. Recent analyses estimate a value of $\beta=0.44_{-0.04}^{+0.05}$ for the total population NSXBs, with the atoll sub-population (which SAX J1810 is likely a member of) having $\beta=0.71_{-0.09}^{+0.11}$ \citep{2018MNRAS.478L.132G}. Both values of $\beta$ reject our measurements at the ${>}\,3\sigma$ level. Therefore, the observed radio emission likely originates from two components, with the most likely candidates of the persistent emission being either a discrete jet ejection or an unrelated, spatially coincident source.

We disfavor an origin due to jet ejection(s). First, the average decay timescale of an ejection event is ${\ll}\,1\,$year, and thus a jet ejection persisting for ${\sim}\,2\,$years and showing no significant decrease in the measured flux density is, in itself, unlikely. Long-lasting jet ejecta have been observed from BHXBs and are thought to be the result of jet-ISM interactions driving in situ particle acceleration and long-term synchrotron emission \citep[e.g.,][]{2005ApJ...632..504C,bright2020,10.1093/mnras/stab864,2023ApJ...948L...7B}. However, such long-lasting ejecta have never been observed in NSXB (likely due to their weaker, lower-luminosity jets being unable to power such long-term emission), and when observed in BHXBs, the radio emission of long-lived ejecta is strongly variable. Second, our VLA follow-up observations suggest a $3\sigma$ upper limit on the radio spectral index of $\alpha\,{<}-1.1$, significantly steeper than expected from optically-thin synchrotron emission from a jet ejection ($\alpha\,{\sim}\,-0.7)$. Lastly, our observations show no evidence of ballistic motion despite the source persisting for ${\sim}\,2\,$years, which would be the strongest evidence for a jet ejecta origin of the persistent emission. If the persistent emission originated from jet ejecta, we would have had to observe a long-lasting, non-variable, spectrally steep ejecta showing no motion on the sky. Therefore, we can rule out a jet ejecta origin with high confidence.   

\begin{figure}
    \centering
    \includegraphics[width=0.9\linewidth]{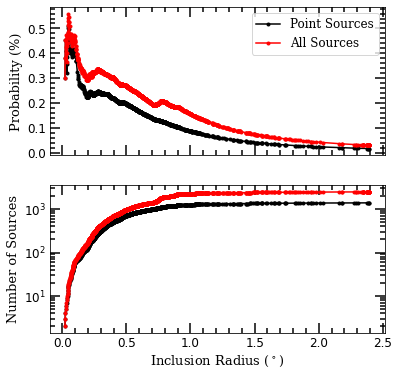}
    \caption{(top panel) The probability of a chance spatial coincidence between SAX J1810 and an unrelated background source as a function of the `inclusion radius'. (bottom panel) The number of sources within the inclusion radius. We include data for both the unresolved (black line) and for unresolved $+$ extended source populations (red line). We note that the sharp increase and peaks close to SAX J1810 correspond to a regime susceptible to low-count statistics. Regardless we adopt the peak of the red curve as the most conservative estimate of the chance coincidence probability.}
    \label{fig:spurious_prob}
\end{figure}

To estimate the probability of a spurious spatial coincidence with an unrelated source in the field we used the Python Blob Detector and Source Finder \citep[PyBDSF; ][]{2015ascl.soft02007M} to make a catalog of all sources (in each image) with a flux density ${>}\,74\,\mu$Jy ($3\sigma$ lower than the average persistent radio flux density). We use the deep 2023 observations as their lower rms noise ($10\,\mu$Jy vs.\ $20\,\mu$Jy in 2021) makes PyBSDF less prone to mistaking spurious noise spikes as real sources. Due to flux variability, each image catalog has a different number of sources. As a result, we conservatively use the 2023 May 22 image as it has more sources than the August observation and, therefore, a larger source density. We calculate the source density and then convert it to the expected number of sources within a $3^{\prime\prime}$ radius. The choice of $3^{\prime\prime}$ was motivated by the scatter of our best-fit positions. Using the expected number of sources, we then calculate the Poissonian probability of a chance coincidence of one or more unrelated background sources. The instrument's sensitivity decreases as a function of radial distance from the phase center of the array, and thus, there is a progressively smaller number of sources cataloged at larger separations from the phase centre (decreasing the source density). We applied a cut when calculating the probability to investigate this potential bias, only including sources within a certain distance from the phase centre in our calculations. In Fig.~$\ref{fig:spurious_prob}$, we show the chance coincidence probability as a function of the aforementioned `inclusion radius' for only unresolved sources (following the criteria from Appendix \ref{sec:appendix_radio_astronomy}) and for both unresolved and extended sources (all sources). We adopt the peak value for all sources as our conservative estimate of the chance coincidence probability (i.e., ${\sim}\,0.6\%$).  

Radio-bright active galactic nuclei (AGN) are the dominant population of unresolved background sources. However, background AGN have an average spectral index of $\alpha\,{\sim}-0.7$. We use two recent surveys of background AGN spectral indexes to estimate the probability of finding a steep spectrum AGN. \citet{randall2012} calculated the spectral index of 166 AGN using 325, 610, and 1400$\,$MHz flux densities. Only 43 sources had an $\alpha\,{<}-1.1$ corresponding to a probability of ${\sim}\,26\%$. In a more recent, larger sample size survey, \citet{Gasperin2018} measured the spectral indexes of ${\sim}\,540000$ radio sources (using 147 and 1400$\,$MHz flux densities), with only a subset of ${\sim}\,32000$ having an appropriately steep $\alpha$. The corresponding probability is ${\sim}\,6\%$. Adopting the older catalog probability as a conservative estimate, we calculate the total probability of finding a spurious radio AGN with a sufficiently steep spectral index as ${\sim}\,0.16\%$ (a ${\sim}\,3.2\sigma$ event). Alternatively, the spectral index could suggest an origin from a class of sources known to have steep spectral indexes. The most common steep spectrum source is pulsars, with average spectral indexes of ${\sim}-1.6$ \citep{2018MNRAS.473.4436J}. We searched the Australian Telescope National Facility pulsar catalog \citep{2005AJ....129.1993M} for any nearby known radio pulsars but found no pulsars within a radius of $0.6^\circ$. Given that there are only 3000 known radio pulsars (corresponding to an expectation value of ${\sim}\,2\times10^{-7}$ pulsars within a $3^{\prime\prime}$ radius), there is a chance coincidence probability of ${\sim}\,0.002\%$. When considering that pulsars tend to be distributed in the Galactic plane (${\sim}\,20\%$ of the sky), and SAX J1810 is also in the galactic plane, the chance coincidence probability would increase by a factor of ${\sim}\,5$ but is still less likely than the AGN scenario. We note that the persistent emission would correspond to a time-averaged flux of a pulsar; as a result, recent surveys that looked at this part of the sky would have detected a pulsed source \citep[e.g.,][]{keith2010}. Moreover, MeerKAT's pulsar timing backend \citep[i.e., MeerTRAP][]{sanidas2018} was operational during all of our observations but did not detect any pulsed emission from the source. Therefore, our estimated coincidence probability between SAX J1810 and an unknown pulsar is most likely an overestimate.   

There is a small possibility that the persistent radio-emission is local to SAX J1810. Transitional millisecond pulsars (tMSPs) --- accreting neutron stars that transition between accretion-powered (i.e., NSXB-like) and radio pulsar behaviour --- have shown anomalously bright radio emission while actively accreting. For instance, the tMSP, 3FGL J0427.9$-$6704, was measured at a point on the $L_R$--$L_X$ relation that was also more consistent with the population of black hole X-ray binaries; however, its X-ray luminosities were a factor of ${\gtrsim}\,3$\, larger than our upper limits on MJD 60086 \citep[e.g.,][]{Li2020}. Other tMSPs (i.e., PSR J1023+0038) have even exhibited anti-correlations between radio and X-ray luminosities, which could allow for bright radio emission absent any X-ray detections  \citep{Bogdanov2018}.

However, the properties of SAX J1810 are inconsistent with what is expected from tMSPs. Firstly, SAX J1810 does not show radio pulsations during X-ray quiescence \citep[although eclipses or highly compact, elliptical binary orbits can prevent the detection of pulsations from tMSPs][]{lorimer2004, Papitto2013}. Second, at X-ray luminosities $\leq\,10^{33}{\rm\,erg\,s^{-1}}$, tMSPs spectra are non-thermal \cite[$\Gamma\,{\leq}\,1.7$][]{Linares2014,Bogdanov2018,Li2020}, whereas SAX J1810 is thermally dominated \citep[$\Gamma\,{\geq}\,3$][]{2004MNRAS.349...94J, Allen2018}. Lastly, SAX J1810 does not exhibit any of the rapid X-ray variability that results from switching between different accretion modes (during outburst), showing, at most, modest variability \citep{Allen2018}. Although it cannot be conclusively ruled out, we find it unlikely that the persistent radio emission results from SAX J1810 being a tMSP.

Local emission, tMSP or otherwise, is difficult to reconcile with the variability in the position, as the source is spatially unresolved. Using the scatter in the measured position (${\sim}\,3^{\prime\prime}$) as a proxy for the expected separation of the two-source scenarios (i.e., the persistent emission is non-local), then observations by an instrument with sufficient angular resolution and sensitivity  (e.g., the VLA in A-configuration or the Square Kilometer Array) during future outbursts when the compact jet is ‘on’ should be able to spatially resolve two distinct components. If only a single source is observed, and there continues to be temporally correlated evolution in the radio/X-ray light curves, this would strongly support the scenario where the persistent radio emission is local to SAX J1810.

\section{Summary and Conclusions}
\label{sec:conclusions}

We have presented our ${\sim}\,2\,$year joint radio and X-ray monitoring of the neutron star X-ray binary SAX J1810.8$-$2609. Our observations include dense (i.e., weekly cadence) observations during the source's 2021 outburst and a collection of late-time observations in 2023. The X-ray spectral properties suggested that the source remained in the hard state throughout the entire 2021 outburst. Moreover, the radio and X-ray luminosities show a temporally correlated evolution, characteristic of a hard state radio jet. We discovered a spatially coincident, persistent steep-spectrum radio source that shows no correlation with the simultaneous X-ray flux. Therefore, during the outburst, the radio emission originated from a superposition of two components: a variable hard state compact jet (${\lesssim}\,100\mu$Jy), and the unknown persistent source (${\sim}\,90\mu$Jy). The spectral index and evolution of the persistent source are inconsistent with jet ejecta. We conservatively estimated the probability of a chance coincidence with an unrelated spectrally steep background source, and although low (${\sim}\,0.16\%$), a background AGN seems to be the most plausible scenario. 

SAX J1810.8$-$2609 is known to go into outburst every ${\sim}\,5\,$years, and future outbursts should focus on identifying the source of the persistent emission. Of the current generation of radio telescopes, the VLA (A-configuration) and the Very Long Baseline Array (VLBA) both have sufficient angular resolution and sensitivity to resolve two ${\sim}\,100\,\mu$Jy sources (assuming a separation of ${\sim}\,3^{\prime\prime}$). Moreover, next-generation radio interferometers, such as the Square Kilometer Array (SKA; of which MeerKAT is a pathfinder), would be able to reach the desired sensitivity with a fraction of the observing time \citep[i.e., ${\sim}\,10\,\mu$Jy rms for ${\lesssim}\,3\,$minutes on source;][]{braun2019}. During the next outburst, if a second unrelated source is ruled out, follow-up observations should focus on understanding what physical mechanism is driving the persistent radio emission, whether the source is a tMSP or otherwise.

\section*{Acknowledgements}

We extend our sincere thanks to all of the NRAO, SARAO, and \textit{Swift}-XRT staff involved in the scheduling and execution of these observations. We thank Kaustubh Rajwade for useful discussions on the completeness of pulsar catalogues. We thank Ben Stappers for searching for pulsed emission in the MeerTRAP observations. We thank Craig Heinke for useful discussions on the X-ray properties of transitional millisecond pulsars. Finally, we thank the referee for their insightful and helpful comments. 

The MeerKAT telescope is operated by the South African Radio Astronomy Observatory, which is a facility of the National Research Foundation, an agency of the Department of Science and Innovation. We acknowledge the use of public data from the \textit{Swift} data archive. This research has made use of MAXI data provided by RIKEN, JAXA and the MAXI team. The National Radio Astronomy Observatory is a facility of the National Science Foundation operated under cooperative agreement by Associated Universities, Inc.

AKH and GRS are supported by NSERC Discovery Grant RGPIN-2021-0400. JvdE acknowledges a Warwick Astrophysics prize post-doctoral fellowship made possible thanks to a generous philanthropic donation.

AKH and GRS respectfully acknowledge that they perform the majority of their research from Treaty 6 territory, a traditional gathering place for diverse Indigenous peoples, including the Cree, Blackfoot, Métis, Nakota Sioux, Iroquois, Dene, Ojibway/ Saulteaux/Anishinaabe, Inuit, and many others whose histories, languages, and cultures continue to influence our vibrant community.

\section*{Data Availability}

Data from MeerKAT are available through the SARAO data archive: \url{https://apps.sarao.ac.za/katpaws/archive-search}. Data from the VLA are available through the VLA data archive (Project ID 23A--417): \url{https://data.nrao.edu/portal}. Data from the \textit{Swift}-XRT are publicly available through the \textit{Swift} archive: \url{https://www.swift.ac.uk/swift_portal}. The authors make their flagging and calibration scripts, imaging results, and analyses available at: \url{https://github.com/AKHughes1994/SAXJ1810_2023}. The astrometry routine is available at: \url{https://github.com/AKHughes1994/AstKAT}. 



\bibliographystyle{mnras}
\bibliography{main} 




\appendix
\section{Radio Astrometry}
\label{sec:appendix_radio_astronomy}
Our observations constitute the first radio detections of SAX J1810, and therefore, we designed a novel astrometric routine to test whether the radio emission is spatially coincident with the archival X-ray position of 18:10:44.47 $-$26:09:01.2 \citep[with its 0.6 arcsec error;][]{2004MNRAS.349...94J}. We divided our astrometric analysis into two components; the first measures the random inter-epoch variability of each source position, quantifying the effects of noise fluctuations (\textit{relative} astrometry), and the second measures the global offsets due to systematic effects in the instrumentation (\textit{absolute} astrometry). The following section outlines our astrometry routine. 

For unresolved sources (i.e., point sources) in synthesis radio images, the relative astrometric error is most often determined by the centroiding accuracy of the Gaussian fitting following deconvolution routines. As the shape of a point source adopts the shape of the synthesized beam in the absence of noise, the astrometric precision decreases with an increasing beam size. The error on the relative astrometry is often described as a function of two components: a signal-to-noise (SNR) dependency and a lower limit set by a systematic threshold. The most commonly assumed signal-to-noise scalings are, 1/SNR, or 1/($2\cdot\text{SNR}$). The systematic threshold is assumed to be some fraction of the synthesized beam size. A common assumption is a lower limit of $10\%$ of the synthesized beam size (e.g., for standard observing with the VLA\footnote{see here; \url{https://science.nrao.edu/facilities/vla/docs/manuals/oss/performance/positional-accuracy}}). We define a generalized (relative) astrometric error with the following functional form, 

\begin{align}
    \sigma = \sqrt{(A\cdot\text{SNR})^2 + B^2},
\end{align}
where $\sigma$ is the relative astrometric error expressed in units of synthesized-beam full widths at half-maxima (FWHM); and $A$ and $B$ are dimensionless variables that describe the SNR scaling and systematic threshold, respectively. Using PyBDSF, we generated a catalogue of (elliptical Gaussian) sources in each image; our parameters of interest were the right ascension (RA), declination (Dec), major axis FWHM of the source, minor axis FWHM of the source, peak flux density ($F_p$), total island flux density\footnote{PyBDSF groups sources into islands, where an island is defined as a continuous region of pixels with a flux value above a user-defined threshold and at least one pixel has a flux larger than a higher (also user-defined) threshold. For large islands (i.e., extended emission), PyBDSF will fit multiple sources to a single island. For our fitting, we used $3\sigma$ and $4\sigma$ for our thresholding.} ($F_i$), and local rms. As SAX J1810 is isolated and unresolved, we trimmed the PyBDSF catalogue to include only similarly unresolved and isolated sources. We defined a source as unresolved if the source FWHMs deviated by ${\leq}\,25\%$ from the synthesized beam shape. Similarly, a source is classified as isolated if the peak flux is within $\,25\%$ of the island flux (e.g., $|F_p/F_i - 1|\,{\leq}\,0.25$). Our routine calculates the average signal-to-noise of each source in the catalogue, and, therefore, we exclude bright transients and strongly variable sources, as their SNR ratio will vary drastically epoch-to-epoch. A source is classified as transient/variable and omitted from the sample if the source is missing from ${>}\,25\%$ of the epochs or has a maximum and minimum flux density separated by a factor ${\geq}\,2$. Lastly, to mitigate biasing from poor far-field calibration errors (e.g., from antenna pointing errors), we fit the sources that are within the inner ${\sim}\,50\%$ of the primary beam FWHM (i.e., sources within 0.3$^\circ$ of the phase centre).

\renewcommand{\arraystretch}{1.1}
\begin{table}
    \centering
    \begin{tabular}{cccccc}
  \Xhline{3\arrayrulewidth}
    Fit Type$^a$ & Dir.  & $A$ ($\%$) & $B$ ($\%$)$^b$ & Pop.$^c$ & $\chi^2$(dof) \\
      \Xhline{3\arrayrulewidth}
\multirow[t]{4}{*}{Uncorrected} & \multirow[t]{2}{*}{RA} & \multirow[t]{2}{*}{$50.0_{-1.0}^{+1.0}$} & \multirow[t]{2}{*}{$1.69_{-0.05}^{+0.05}$} & 1 & 159(120) \\
 &  &  &  & 2 & 1189/556 \\ 
 & \multirow[t]{2}{*}{DEC} & \multirow[t]{2}{*}{$46.1_{-0.9}^{+0.9}$} & \multirow[t]{2}{*}{$1.39_{-0.05}^{+0.05}$} & 1 & 164(120) \\
&  & & & 2 & 1010/556 \\ 
\hline 
\multirow[t]{4}{*}{Corrected} & \multirow[t]{2}{*}{RA} & \multirow[t]{2}{*}{$47.3_{-0.7}^{+0.7}$} & \multirow[t]{2}{*}{$0.39_{-0.03}^{+0.03}$} & 1 & 287(120) \\
 &  &  &  & 2 & 3084(556) \\ 
 & \multirow[t]{2}{*}{DEC} & \multirow[t]{2}{*}{$47.6_{-0.9}^{+0.9}$} & \multirow[t]{2}{*}{$0.18_{-0.04}^{+0.04}$} & 1 & 159(120) \\
&  & & & 2 & 2273(556) \\ 

\Xhline{3\arrayrulewidth}
    \end{tabular}
\caption{Relative Astrometry Parameters: $^a$ This column indicates whether the fitting omitted (uncorrected) or used (corrected) the epoch-to-epoch astrometry correction; $^b$ The fitting parameters $A$ and $B$ are expressed as a fraction of the synthesized beam FWHM for both the RA and Dec directions; $^c$ This column indicates the population of sources used for the corresponding $\chi^2$ calculations. Population 1 is the nearby (${<}\,0.3^\circ$) isolated point sources used in the fittings. Population 2 includes all isolated point sources, regardless of distance from the phase centre. The contrast between the $\chi^2$ values of Population 1 and 2 highlights the effects of far-field errors.}
\label{tab:relative_astrometry}
\end{table}

As the MeerKAT synthesized beam is an elliptical Gaussian, we solve for $A$ and $B$ independently along the RA and Dec directions. Below, we outline our fitting routine:
\begin{enumerate}
    \item For each source, calculate an average SNR and an average position. Calculate the RA/Dec offset from the average position for every source in each epoch using the average position.
    \item Estimate the error in the astrometric precision of each source by bootstrapping the offsets, adopting the median value of the bootstrapped sample as an initial guess for $\sigma$ and the ranges between the median and the $15^\text{th}$/ $85^\text{th}$ percentiles as the 1$\sigma$ $(-)/(+)$ uncertainties ($\Delta_\sigma$).
    \item Using the $\sigma$ estimates and the average SNR, solve for the scaling parameters $A$ and $B$ (i.e., the \textit{uncorrected} fit). The fit implements an MCMC routine and follows the same approach detailed in \ref{sec:methods_xray_burst}.
    \item Solve for the (inverse-variance weighted) average offset of all sources in each epoch (i.e., the \textit{epoch-to-epoch correction}) weighting each offset using the uncorrected fit.
    \item Correct the source offsets with the epoch-to-epoch correction and re-solve for $A$ and $B$ with the updated -- corrected -- offsets.
    \item Repeat (ii)$\rightarrow$(v) until the fitting converges on solutions for $A$ and $B$. We defined a convergence parameter $C = (\sigma_i - \sigma_{i-1})/\Delta_{\sigma}$; i.e., the difference between the astrometric error of a source for the current ($i$) and previous ($i-1$) iterations in units of $\Delta_\sigma$. The fit is said to have converged after three consecutive iterations with a mean value of $C\,{<}\,0.1$. The post-convergence fit is the \textit{corrected} fit. Record the final epoch corrections. 
\end{enumerate}

\begin{figure*}
    \centering
    \includegraphics[width=0.9\linewidth]{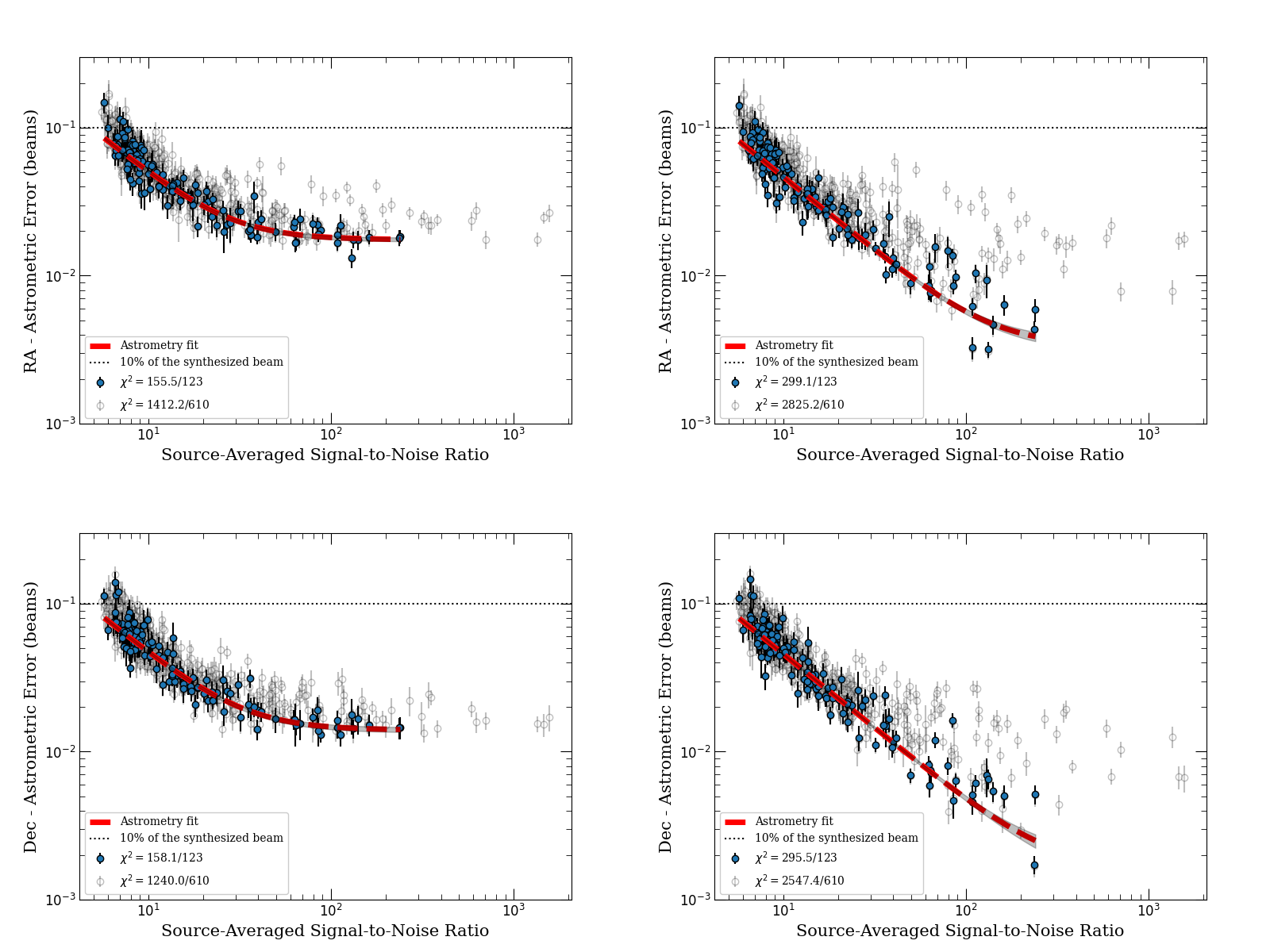}
    \caption{The relative astrometric fits for the population of isolated point-like sources: (top left) uncorrected Dec; (top right) corrected Dec; (bottom left) uncorrected RA; (bottom right) corrected RA. The sources used for fitting (i.e., 125 sources within 0.3$^\circ$ of the phase center) are given by the solid blue circles, and the total population of isolated point-like sources (612 sources) is shown as hollow black circles. The uncorrected fits are marginally acceptable for the fitted population (reduced $\chi^2\,{\sim}\,1.3$ for 123 degrees of freedom), although the corrected fits are poor (reduced $\chi^2\,{>}\,2.0$ for 610 degrees of freedom). Furthermore, the fits (uncorrected and corrected) are poor matches to the total population of isolated point-like sources, suggesting that far-field effects (especially at high signal-to-noise ratios) are significant. Overall, the fits show that the systematic limit is well below $10\%$ of the synthesized beam and that at SNR$\,{>}\,20$, the global epoch affects (i.e., affecting every source in a given epoch) are the dominant astrometric error.} 
    \label{fig:relative_astrometry}
\end{figure*}

The relative astrometric fitting is shown in Fig.~\ref{fig:relative_astrometry} and the best-fit parameters are tabulated in Table \ref{tab:relative_astrometry}. The uncorrected fits have reduced $\chi^2$ values of ${\sim}\,1.3$ (123 degrees of freedom) in both RA and Dec. Applying the epoch-to-epoch corrections (i.e., the corrected fit) shows a significant worsening of the fit quality with a reduced $\chi^2\,{>}\,2$, suggesting that a single per-epoch correction is not accurately capturing the time-dependent systematics in our observations, and a more complex epoch correction may be appropriate (e.g., one that accounts for distance and direction with respect to the phase center). We intend to expand upon this preliminary work to investigate whether the relative astrometric error is similar across a range of ThunderKAT fields. 

\begin{figure}
    \centering
    \includegraphics[width=0.9\linewidth]{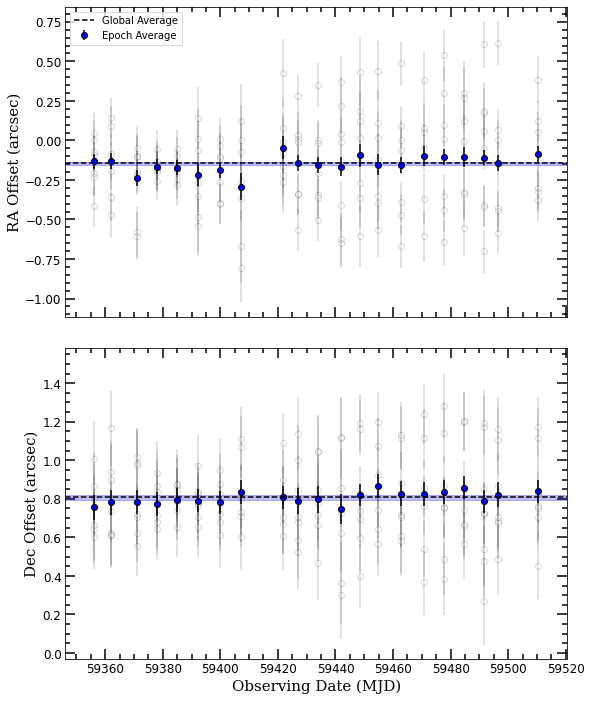}
    \caption{The absolute astrometric corrections from the very long baseline interferometry calibrators in the SAX J1810 field-of-view. The open black circles are the offsets of each calibrator in each epoch. The closed blue circles are the average offset in each epoch. The dashed black line is the time-independent average offset across all sources and all epochs. The blue region shows the 1$\sigma$ errors on the average offset. The per-epoch RA (Dec) average offsets are consistent with the time-independent value at a reduced $\chi^2$ of ${\sim}\,0.65$ (${\sim}\,0.24$) for 22 degrees of freedom. These low $\chi^2$ values suggest that we may be overestimating the relative astrometric error.}  
    \label{fig:vlbi_astro}
\end{figure}

The fits show that (for MeerKAT), the systematic threshold of the relative error is significantly lower than the commonly assumed limit of $10\%$ the size of the synthesized beam. Moreover, the signal-to-noise dependency is similar to the commonly assumed 1/($2\cdot\text{SNR}$) scaling. Due to the residual issues in our modeling, for our SAX J1810 analysis, we conservatively rounded our uncorrected fit values, adopting $A=0.5$  and $B=0.02$ to quantify the relative astrometric errors. 

To correct for absolute astrometry effects, we identified nine sources\footnote{\url{http://astrogeo.org/calib/search.html}} within our field of view that are used as phase calibrators for very long baseline interferometry (i.e., with positions measured at ${<}\,10\,$milliarcsecond precision). Eight of the nine sources met our unresolved and isolated requirement, and we used this sub-sample for absolute astrometric corrections. After applying the epoch-to-epoch correction from the relative astrometric fitting, we measured the offsets of the eight calibrators with respect to their known positions. We then calculate each epoch's weighted mean (weighting each source by their relative astrometric errors). Lastly, we calculated a single time-independent absolute astrometric correction (see Fig.~\ref{fig:vlbi_astro}). The epoch-to-epoch correction removed any (substantial) temporal variability, and, as a result, the per-epoch average offsets are consistent with a single (time-independent) RA/Dec offset. 

The final astrometric error ($\sigma_\text{tot}$) was calculated by adding (in quadrature) the relative astrometric precision ($\sigma$), the error on the epoch-correction ($\sigma_\text{epoch}$), and the error on the absolute offset ($\sigma_\text{abs}$),
\begin{align}
    \sigma_\text{tot} = \sqrt{\sigma^2 + \sigma_\text{epoch}^2 + \sigma_\text{abs}^2}.
\end{align}
These are the errors shown in Fig.~\ref{fig:sax_astro}. We note that given the signal-to-noise ratio of our SAX J1810 detections (SNR$\,{\lesssim}\,$10), the relative astrometry term, $\sigma$, dominates the quoted errors.

\section{Type I X-Ray Burst}
\label{sec:appendix_xray_burst}
\begin{figure*}
    \centering
    \includegraphics[width=0.9\linewidth]{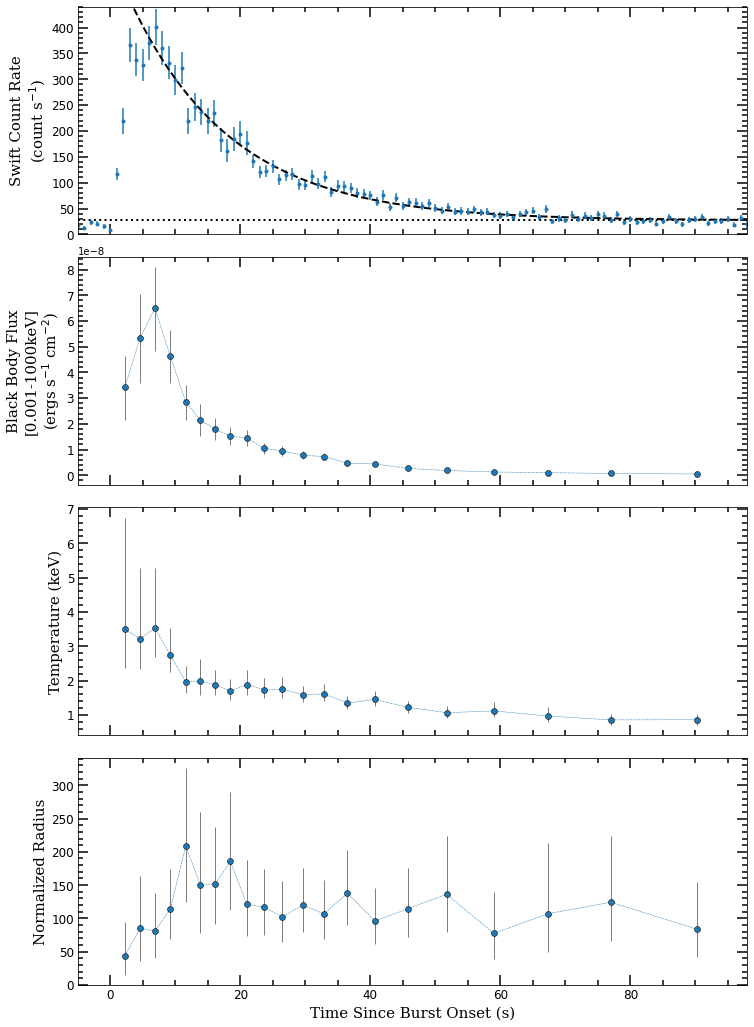}
    \caption{Spectral and timing fits from the Type I X-ray burst observed on 2022 August 7. The top panel shows the 1s-binned fight curves. We overlayed the best-fit exponential decay (dashed line) and the constant (pre-burst) count rate(dotted line); the timing fit parameters are tabulated in Table \ref{tab:timing_fit}. The bolometric X-ray flux (top panel), temperature (2nd panel), and normalized radius (bottom panel) of the black body component do not show conclusive evidence of PRE. }
    \label{fig:xray_burst}
\end{figure*}

Figure \ref{fig:xray_burst} shows the parameters of the 2022 August 7 (MJD 59433) Type I X-ray burst. The top panel shows the 1s-binned light curves and the timing fits; the second panel shows the bolometric X-ray flux of the blackbody component; the third panel shows the temperature of the blackbody component; and the bottom panel shows the normalized radius of the blackbody component, defined as $R^2/D^2$, where $R$ is the source radius in units of km and $D$ is the distance to the source in units of 10$\,$kpc. 

\begin{table}
    \centering
    \begin{tabular}{cccc}
  \Xhline{3\arrayrulewidth}
    $\tau$\,(s) & $R_0{\rm\,(counts\,s^{-1})}$ & $A{\rm\,(counts\,s^{-1})}$ & $\chi^2$(dof) \\
      \Xhline{3\arrayrulewidth}

    $15.8\,{\pm}\,0.2\,$ & $27.0\,{\pm}\,0.5\,$ & $332\,{\pm}\,3\,$ & 86(89) \\ 

\Xhline{3\arrayrulewidth}

    \end{tabular}
    \caption{X-ray burst timing parameters.}
    \label{tab:timing_fit}
\end{table}

The burst began its rise at 14:14:12 on 2021 August 7 (MJD 59433.59319), reaching a peak count rate of ${\sim}\,400$\counts with a rapid $7\,{\pm}\,1\,$s rise time before decaying for the remainder of our observations. The timing fit converged on an \textit{e}-folding decay time of $\tau = 15.8\,{\pm}\,0.2\,$s (full fit parameters in Table \ref{tab:timing_fit}). The burst parameters are consistent with the MINBAR burst catalog \cite{2020ApJS..249...32G} in both rise ($3.4_{-2.4}^{+5.6}\,$s) and \textit{e}-folding decay times ($8_{-4}^{+21}\,$s). 

During its 2007 outburst, SAX J1810 exhibited 531.8$\,$Hz oscillations in the light curves of a Type I X-ray burst, likely the result of the spin frequency of the neutron star \citep{2018ApJ...862L...4B}. Following the prescription outlined in \citet{2018ApJ...862L...4B} we searched for burst oscillations in our (1.8$\,$ms resolution) light curves by calculating the power spectrum in sliding windows with widths of 0.5, 1, 2, and 4$\,$s, where each subsequent window is offset by 0.5$\,s$ from the previous one. We found no evidence of burst oscillations. However, the temporal resolution of \textit{Swift}-XRT WT mode ($1.8\,$ms) makes our power spectra insensitive to frequencies above ${\sim}\,280\,$Hz. Assuming the oscillations result from the spin period of the neutron star, we do not expect the oscillation frequency to evolve drastically between the 2007 and 2021 outbursts. 

Furthermore, SAX J1810 is known to exhibit PRE \citep[i.e., during the 1998 outburst a PRE signature provided the current distance constraint of $4.9\,{\pm}\,0.3\,$kpc; ][]{2000ApJ...536..891N}. Therefore, we performed time-resolved intra-epoch spectral modelling to search for evidence of PRE. We observe some evolution of the radius and temperature, although the large errors greatly reduce their significance. Assuming a distance of $4.9\,$kpc, the radius of the blackbody component ranges from $3.2_{-2.6}^{+3.5}$ to $6.7_{-4.2}^{+5.0}\,$km (i.e., from ${\sim}\,5$-to-$100\%$ of the neutron stars surface assuming a 10$\,$km stellar radius). However, the evolution of the radius and temperature does not occur alongside a period of (approximately) constant X-ray flux; thus, we do not detect PRE. 

\section{Data Tables}
\renewcommand{\arraystretch}{1.5}
\begin{table*}
  \centering
  \caption{Radio properties of SAX J1810}
  \label{tab:radio_monitoring}  
  \begin{tabular}{cccccc}
  \Xhline{3\arrayrulewidth}
    MJD  & Date & Instrument & Central Frequency [GHz] & Radio Flux Density $[\mu{\rm Jy}]$ \\
  \Xhline{3\arrayrulewidth}
59356 & 2021-05-22 & MeerKAT & 1.3 & $232\pm18$ \\
59362 & 2021-05-27 & MeerKAT & 1.3 & $197\pm18$ \\
59371 & 2021-06-05 & MeerKAT & 1.3 & $207\pm17$ \\
59378 & 2021-06-12 & MeerKAT & 1.3 & $143\pm15$ \\
59385 & 2021-06-19 & MeerKAT & 1.3 & $128\pm18$ \\
59392 & 2021-06-27 & MeerKAT & 1.3 & $139\pm20$ \\
59400 & 2021-07-04 & MeerKAT & 1.3 & $135\pm16$ \\
59407 & 2021-07-12 & MeerKAT & 1.3 & $153\pm26$ \\
59422 & 2021-07-26 & MeerKAT & 1.3 & $175\pm20$ \\
59427 & 2021-07-31 & MeerKAT & 1.3 & $197\pm17$ \\
59434 & 2021-08-07 & MeerKAT & 1.3 & $140\pm19$ \\
59442 & 2021-08-15 & MeerKAT & 1.3 & $163\pm24$ \\
59449 & 2021-08-22 & MeerKAT & 1.3 & $110\pm22$ \\
59455 & 2021-08-28 & MeerKAT & 1.3 & $140\pm23$ \\
59463 & 2021-09-05 & MeerKAT & 1.3 & $97\pm17$ \\
59471 & 2021-09-13 & MeerKAT & 1.3 & $98\pm22$ \\
59478 & 2021-09-20 & MeerKAT & 1.3 & $83\pm17$ \\
59485 & 2021-09-27 & MeerKAT & 1.3 & $142\pm26$ \\
59492 & 2021-10-04 & MeerKAT & 1.3 & $92\pm17$ \\
59497 & 2021-10-09 & MeerKAT & 1.3 & $97\pm17$ \\
59511 & 2021-10-23 & MeerKAT & 1.3 & $82\pm16$ \\
60086 & 2023-05-22 & MeerKAT & 1.3 & $112\pm12$ \\ 
60142 & 2023-07-17 & VLA & 3.0 & ${<}30$ \\ 
60142 & 2023-07-17 & VLA & 6.0 & ${<}18$ \\ 
60169 & 2023-08-13 & MeerKAT & 1.3 & $75\pm11$ \\ 

\Xhline{3\arrayrulewidth}
  \end{tabular}
\end{table*}

\renewcommand{\arraystretch}{1.5}
\begin{table*}
  \centering
  \caption{X-ray spectral properties from our \textit{Swift}-XRT monitoring of SAX J1810. The table includes the following model parameters: power law photon index ($\Gamma$), power law flux ($F_\text{PL}$), black body temperature ($kT$), black body flux ($F_\text{BB}$), and the total X-ray flux ($F_\text{tot}$).}
  \label{tab:swiftxrt_params}  
  \begin{tabular}{ccccccccc}
  \Xhline{3\arrayrulewidth}
    MJD  & Date & $E$ & $\Gamma$ & $F_\text{PL}$ & $kT$ & $F_\text{BB}$ & $F_\text{tot}$ & $\chi^2$(dof) \\    
      &  & (keV) & & $[10^{-11}\,{\rm erg\,s^{-1}\,cm^{-2}}]$ & (keV) & $[10^{-11}\, {\rm erg\,s^{-1}\,cm^{-2}}]$ & $[10^{-11}\,{\rm erg\,s^{-1}\,cm^{-2}}]$ &  \\    
  \Xhline{3\arrayrulewidth}
\multirow[t]{2}{*}{59364} & \multirow[t]{2}{*}{2021-05-30} & 0.5--10 & ${1.42}_{-0.08}^{+0.09}$ & ${131.9}_{-8.2}^{+8.5}$ & ${0.68}_{-0.07}^{+0.09}$ & ${23.6}_{-2.7}^{+2.0}$ & ${155.5}_{-8.6}^{+8.8}$ & 257(278) \\
& & 1.0--10 & ${1.42}_{-0.04}^{+0.08}$ & ${118.0}_{-7.5}^{+7.6}$ & ${0.68}_{-0.07}^{+0.09}$ & ${21.7}_{-2.5}^{+1.9}$ & ${139.7}_{-7.9}^{+7.9}$ & 257(278) \\
\hline
\multirow[t]{2}{*}{59378} & \multirow[t]{2}{*}{2021-06-13} & 0.5--10 & ${1.82}_{-0.11}^{+0.16}$ & ${24.0}_{-2.7}^{+2.7}$ & ${0.76}_{-0.07}^{+0.08}$ & ${9.0}_{-0.9}^{+0.8}$ & ${33.0}_{-2.8}^{+2.8}$ & 191(190) \\
& & 1.0--10 & ${1.82}_{-0.11}^{+0.13}$ & ${19.5}_{-1.4}^{+2.5}$ & ${0.76}_{-0.11}^{+0.08}$ & ${8.5}_{-0.9}^{+0.8}$ & ${28.0}_{-1.6}^{+2.7}$ & 191(190) \\
\hline
\multirow[t]{2}{*}{59385} & \multirow[t]{2}{*}{2021-06-20} & 0.5--10 & ${2.88}_{-0.08}^{+0.18}$ & ${26.6}_{-1.5}^{+1.7}$ & ${1.16}_{-0.09}^{+0.10}$ & ${12.7}_{-0.8}^{+0.7}$ & ${39.3}_{-1.7}^{+1.8}$ & 217(186) \\
& & 1.0--10 & ${2.89}_{-0.18}^{+0.18}$ & ${13.5}_{-1.7}^{+2.0}$ & ${1.16}_{-0.08}^{+0.10}$ & ${12.5}_{-0.4}^{+0.7}$ & ${26.0}_{-1.8}^{+2.1}$ & 216(186) \\
\hline
\multirow[t]{2}{*}{59392} & \multirow[t]{2}{*}{2021-06-27} & 0.5--10 & ${1.20}_{-0.14}^{+0.08}$ & ${24.9}_{-2.2}^{+2.4}$ & ${0.64}_{-0.03}^{+0.04}$ & ${14.1}_{-0.7}^{+0.6}$ & ${39.0}_{-2.3}^{+2.4}$ & 237(244) \\
& & 1.0--10 & ${1.22}_{-0.15}^{+0.12}$ & ${23.0}_{-2.0}^{+2.0}$ & ${0.65}_{-0.03}^{+0.03}$ & ${12.9}_{-0.7}^{+0.5}$ & ${35.9}_{-2.1}^{+2.1}$ & 237(244) \\
\hline
\multirow[t]{2}{*}{59399} & \multirow[t]{2}{*}{2021-07-04} & 0.5--10 & ${1.83}_{-0.08}^{+0.10}$ & ${49.6}_{-1.8}^{+3.8}$ & ${0.72}_{-0.13}^{+0.16}$ & ${6.7}_{-1.3}^{+1.0}$ & ${56.2}_{-2.2}^{+3.9}$ & 238(231) \\
& & 1.0--10 & ${1.83}_{-0.04}^{+0.10}$ & ${40.3}_{-3.8}^{+3.4}$ & ${0.73}_{-0.13}^{+0.16}$ & ${6.2}_{-1.3}^{+1.0}$ & ${46.5}_{-4.0}^{+3.5}$ & 238(231) \\
\hline
\multirow[t]{2}{*}{59406} & \multirow[t]{2}{*}{2021-07-11} & 0.5--10 & ${1.61}_{-0.14}^{+0.16}$ & ${30.0}_{-4.0}^{+3.8}$ & ${0.72}_{-0.11}^{+0.12}$ & ${8.9}_{-1.3}^{+1.0}$ & ${39.0}_{-4.2}^{+4.0}$ & 95(105) \\
& & 1.0--10 & ${1.62}_{-0.14}^{+0.15}$ & ${25.8}_{-3.9}^{+3.6}$ & ${0.72}_{-0.10}^{+0.12}$ & ${8.4}_{-1.3}^{+1.0}$ & ${34.1}_{-4.1}^{+3.7}$ & 95(105) \\
\hline
\multirow[t]{2}{*}{59413} & \multirow[t]{2}{*}{2021-07-18} & 0.5--10 & ${1.35}_{-0.11}^{+0.09}$ & ${46.4}_{-4.8}^{+4.7}$ & ${0.79}_{-0.08}^{+0.11}$ & ${14.9}_{-1.4}^{+1.2}$ & ${61.3}_{-5.0}^{+4.9}$ & 237(194) \\
& & 1.0--10 & ${1.35}_{-0.06}^{+0.12}$ & ${42.0}_{-4.8}^{+3.5}$ & ${0.80}_{-0.09}^{+0.11}$ & ${14.2}_{-1.4}^{+1.3}$ & ${56.2}_{-5.0}^{+3.7}$ & 237(194) \\
\hline
\multirow[t]{2}{*}{59433} & \multirow[t]{2}{*}{2021-08-07} & 0.5--10 & ${1.65}_{-0.08}^{+0.10}$ & ${77.5}_{-6.3}^{+5.9}$ & ${0.91}_{-0.11}^{+0.13}$ & ${14.6}_{-2.0}^{+1.9}$ & ${92.1}_{-6.6}^{+6.2}$ & 318(340) \\
& & 1.0--10 & ${1.66}_{-0.08}^{+0.09}$ & ${65.8}_{-6.5}^{+5.7}$ & ${0.91}_{-0.11}^{+0.13}$ & ${14.1}_{-2.1}^{+1.8}$ & ${80.0}_{-6.8}^{+6.0}$ & 318(340) \\
\hline
\multirow[t]{2}{*}{59437} & \multirow[t]{2}{*}{2021-08-11} & 0.5--10 & ${1.75}_{-0.06}^{+0.07}$ & ${87.0}_{-5.4}^{+5.2}$ & ${0.82}_{-0.24}^{+0.23}$ & ${5.3}_{-2.4}^{+1.4}$ & ${92.3}_{-5.9}^{+5.4}$ & 291(324) \\
& & 1.0--10 & ${1.75}_{-0.03}^{+0.09}$ & ${72.1}_{-5.9}^{+4.8}$ & ${0.84}_{-0.25}^{+0.33}$ & ${5.1}_{-2.4}^{+1.4}$ & ${77.2}_{-6.3}^{+5.0}$ & 291(324) \\
\hline
\multirow[t]{2}{*}{59451} & \multirow[t]{2}{*}{2021-08-25} & 0.5--10 & ${1.43}_{-0.17}^{+0.16}$ & ${16.8}_{-0.9}^{+1.1}$ & ${0.60}_{-0.04}^{+0.05}$ & ${9.0}_{-0.7}^{+0.5}$ & ${25.8}_{-1.1}^{+1.2}$ & 135(142) \\
& & 1.0--10 & ${1.44}_{-0.18}^{+0.15}$ & ${15.0}_{-1.7}^{+1.4}$ & ${0.60}_{-0.04}^{+0.04}$ & ${8.0}_{-0.6}^{+0.5}$ & ${23.1}_{-1.8}^{+1.4}$ & 135(142) \\
\hline
\multirow[t]{2}{*}{59458} & \multirow[t]{2}{*}{2021-09-01} & 0.5--10 & ${1.34}_{-0.15}^{+0.14}$ & ${20.1}_{-1.9}^{+2.1}$ & ${0.60}_{-0.04}^{+0.05}$ & ${9.6}_{-0.6}^{+0.6}$ & ${29.6}_{-2.0}^{+2.2}$ & 169(181) \\
& & 1.0--10 & ${1.35}_{-0.14}^{+0.13}$ & ${18.2}_{-1.7}^{+1.8}$ & ${0.60}_{-0.04}^{+0.04}$ & ${8.5}_{-0.6}^{+0.5}$ & ${26.7}_{-1.8}^{+1.8}$ & 169(181) \\
\hline
\multirow[t]{2}{*}{59462} & \multirow[t]{2}{*}{2021-09-05} & 0.5--10 & ${0.95}_{-0.71}^{+0.42}$ & ${20.4}_{-5.2}^{+4.6}$ & ${0.62}_{-0.06}^{+0.11}$ & ${11.7}_{-1.9}^{+1.7}$ & ${32.1}_{-5.5}^{+4.9}$ & 50(33) \\
& & 1.0--10 & ${0.92}_{-0.71}^{+0.47}$ & ${19.1}_{-4.5}^{+4.3}$ & ${0.62}_{-0.03}^{+0.12}$ & ${10.9}_{-1.9}^{+1.4}$ & ${30.0}_{-4.9}^{+4.6}$ & 50(33) \\
\hline
\multirow[t]{2}{*}{59468} & \multirow[t]{2}{*}{2021-09-11} & 0.5--10 & ${1.23}_{-0.29}^{+0.23}$ & ${10.8}_{-1.5}^{+1.6}$ & ${0.53}_{-0.02}^{+0.04}$ & ${5.9}_{-0.3}^{+0.5}$ & ${16.8}_{-1.5}^{+1.7}$ & 103(94) \\
& & 1.0--10 & ${1.23}_{-0.26}^{+0.23}$ & ${10.0}_{-1.3}^{+1.3}$ & ${0.53}_{-0.04}^{+0.04}$ & ${5.1}_{-0.5}^{+0.4}$ & ${15.1}_{-1.4}^{+1.4}$ & 103(94) \\
\hline
\multirow[t]{2}{*}{59475} & \multirow[t]{2}{*}{2021-09-18} & 0.5--10 & ${0.85}_{-0.61}^{+0.44}$ & ${9.0}_{-0.8}^{+1.8}$ & ${0.53}_{-0.04}^{+0.04}$ & ${6.7}_{-0.8}^{+0.6}$ & ${15.7}_{-1.1}^{+1.9}$ & 77(69) \\
& & 1.0--10 & ${0.87}_{-0.65}^{+0.43}$ & ${8.6}_{-1.5}^{+1.5}$ & ${0.53}_{-0.04}^{+0.04}$ & ${5.8}_{-0.7}^{+0.6}$ & ${14.4}_{-1.6}^{+1.6}$ & 77(69) \\
\hline
\multirow[t]{2}{*}{59482} & \multirow[t]{2}{*}{2021-09-25} & 0.5--10 & ${1.34}_{-0.28}^{+0.19}$ & ${14.1}_{-1.8}^{+1.5}$ & ${0.56}_{-0.06}^{+0.07}$ & ${5.3}_{-0.7}^{+0.5}$ & ${19.5}_{-2.0}^{+1.6}$ & 59(87) \\
& & 1.0--10 & ${1.34}_{-0.27}^{+0.19}$ & ${12.8}_{-1.6}^{+1.6}$ & ${0.56}_{-0.06}^{+0.05}$ & ${4.7}_{-0.6}^{+0.4}$ & ${17.5}_{-1.7}^{+1.6}$ & 59(87) \\
\hline
\multirow[t]{2}{*}{59489} & \multirow[t]{2}{*}{2021-10-02} & 0.5--10 & ${0.84}_{-0.53}^{+0.39}$ & ${12.7}_{-1.8}^{+1.9}$ & ${0.56}_{-0.03}^{+0.02}$ & ${10.9}_{-1.0}^{+1.1}$ & ${23.5}_{-2.1}^{+2.2}$ & 107(110) \\
& & 1.0--10 & ${0.96}_{-0.65}^{+0.29}$ & ${12.3}_{-1.8}^{+1.7}$ & ${0.57}_{-0.03}^{+0.02}$ & ${9.2}_{-0.7}^{+0.9}$ & ${21.5}_{-1.9}^{+1.9}$ & 107(110) \\
\hline
\multirow[t]{2}{*}{59496} & \multirow[t]{2}{*}{2021-10-09} & 0.5--10 & ${0.43}_{-0.43}^{+0.93}$ & ${4.0}_{-0.8}^{+0.8}$ & ${0.47}_{-0.05}^{+0.05}$ & ${2.8}_{-0.7}^{+0.2}$ & ${6.8}_{-1.0}^{+0.8}$ & 29(35) \\
& & 1.0--10 & ${0.45}_{-0.45}^{+0.93}$ & ${3.9}_{-0.4}^{+0.4}$ & ${0.47}_{-0.05}^{+0.04}$ & ${2.3}_{-0.6}^{+0.2}$ & ${6.2}_{-0.7}^{+0.4}$ & 29(35) \\
\hline
\multirow[t]{2}{*}{59504} & \multirow[t]{2}{*}{2021-10-17} & 0.5--10 & --- & --- & --- & --- & ${0.21}_{-0.12}^{+0.12}$ & 59(95) \\
 &  & 1.0--10 & --- & --- & --- & --- & ${0.18}_{-0.10}^{+0.11}$ & 59(95) \\
\hline
\multirow[t]{2}{*}{59511} & \multirow[t]{2}{*}{2021-10-24} & 0.5--10 & --- & --- & --- & --- & ${0.05}_{-0.03}^{+0.04}$ & 7(7) \\
 &  & 1.0--10 & --- & --- & --- & --- & ${0.04}_{-0.03}^{+0.04}$ & 7(7) \\
\hline

  \Xhline{3\arrayrulewidth}
  \end{tabular}
\end{table*}

\bsp	
\label{lastpage}
\end{document}